\documentstyle[pra,aps,epsf]{revtex}
\newcommand{\BEQ}{\begin{equation}}
\newcommand{\EEQ}{\end{equation}}
\newcommand{\BEA}{\begin{eqnarray}}
\newcommand{\EEA}{\end{eqnarray}}
\newcommand{\dk}{{\frac{{\rm d}^dk}{(2\pi)^d}}}
\newcommand{\tr}{{\rm tr}}
\renewcommand{\d}{{\,{\rm d}}}
\newcommand{\e}{{\,{\rm e}}}
\renewcommand{\a}{{\alpha}}

\newcommand{\ab}{{\alpha\beta}}
\newcommand{\nn}{\nonumber\\}
\begin{document}
\draft
\title{THEORY OF SITE-DISORDERED MAGNETS}
\author{ Th.M. Nieuwenhuizen$^{(1)}$ and C.N.A. van Duin$^{(2)}$}
\address{$^{(1)}$University of Amsterdam, Van der Waals-Zeeman Instituut\\
Valckenierstraat 65-67, 1018 XE Amsterdam, The Netherlands\\
$^{(2)}$Institute Lorentz for Theoretical Physics, Leiden University\\
P.O.B. 9506, 2300 RA Leiden, The Netherlands }

\date{\today}
\maketitle
\begin{abstract}
In realistic spinglasses, such as ${\rm CuMn}$, ${\rm AuFe}$
and ${\rm EuSrS}$, magnetic atoms are located at random positions.
 Their couplings are determined by their relative positions.
For such systems a field theory is formulated.
In certain limits it reduces to
 the Hopfield model,
 the Sherrington-Kirkpatrick model,
and the Viana-Bray model.
The model has a percolation transition, while for RKKY couplings
the ``concentration scaling'' $T_g\sim c$ occurs.

Within the Gaussian approximation
the Ginzburg-Landau expansion is considered in the clusterglass phase,
that is to say, for not too small concentrations.
Near special points,
the prefactor of the cubic term, or the one of the replica-symmetry-
breaking quartic term, may go through zero. Around such points
new spin glass phases are found.

\end{abstract}
\pacs{ 75.10 Nr, 75.30 Fv, 75.50 Lk}
\section{Introduction}
Prototypes of systems exhibiting a spin glass phase are
the impure metals Cu$_{1-c}$Mn$_c$ and Au$_{1-c}$Fe$_c$, and
the insulator Eu$_{1-c}$Sr$_c$S.~\cite{Mydoshboek}~\cite{FischerHertz}
Their common property is frustration of magnetic bonds,
 which arises from the
combination of site disorder and the presence of both
 ferromagnetic and antiferromagnetic exchange constants.
In metals the exchange is the long range, oscillating
 RKKY interaction;
in an insulating SG the interaction is often antiferromagnetic
for nearest neighbors and ferromagnetic for next
nearest neighbors, while vanishing for more distant pairs.

Edwards and Anderson
 proposed to describe the spinglass phase
by spins with fully random couplings \cite{EA}.
The mean field limit of this model for
 Ising spins is the Sherrington-Kirkpatrick (SK)
 model \cite{SK}.
The diluted version, where only a fraction
of the random bonds is present, is
called the Viana-Bray model \cite{VB}.

The mean field structure of the SG-phase was derived by
Parisi and coworkers. There are infinitely many
thermodynamic states, and the
order parameter is a function $q(x)$ or $P(q)$, which
describes the probabilities of overlaps between pairs of states.\cite{MPV}

The calculation of corrections to mean field theory has been
very laborious. For a recent review, see ~\cite{dDKT}.

The question what remains of mean field theory in low dimensions
is still fiercely debated. Nevertheless, the best numerical
simulations done so far are well analyzed within a mean field picture {}~\cite{meanfieldpicture}.

Mean field notions such as branching ratios and ultrametric organization
of states have turned out useful in the interpretation of experiments.
Indeed, noise experiments on mesoscopic metallic spin glasses
have supported the mean field picture \cite{Weissman}.

We are thus at the stage where theoretical results are mainly gained from
random bond systems, while experiments are mostly performed on random site
systems. The understanding of the spin glass phase in random site
systems has long been a challenge for the field. Some time ago
one of the authors formulated a field theoretic approach for this
purpose \cite{NEPL}.
In certain limit close analogies were established
with known spin glass phases in well known models,
 such as the Sherrington-Kirkpatrick model,
the Hopfield model, and the Viana-Bray model.

The first purpose of the present work is to provide the details of that
work. This will be done in section 2. In section 3 one of the
results will be rederived via the Gaussian variational approach.
In section 4 we consider a related model with combined bond- and
site disorder. In section 5 we extend the approach of section 2 to
vector spins. In section 6 we analyze the equations of section 2
for low $T$ and find that replica symmetry must always be broken.

More recently we have considered the Ginzburg-Landau theory
of site-disordered spin glasses~\cite{NvDuin}.
It was argued that the prefactor of
the cubic or most relevant quartic term may vanish at special points
and then change sign. Around such
points new types of spin glass phases occur. In section 7
we shall provide the details of that approach.
In section 8 we close with a summary.

\section{Thermodynamics as a set of coupled order parameters}

We discuss a field theory for a random magnet with site-disorder.
First we consider a system of Ising spins with Hamiltonian
\begin{equation}
\label{Hising}
{\cal H}(s)=-\frac{1}{2}
\sum_{r\,r'}J(r-r')s_rs_{r'}c_rc_{r'}-H\sum_r s_rc_r
\end{equation}
with sums over all $N$ sites of
a regular lattice in $d$ dimensions, with periodic boundary
conditions. The $c_r$ are random occupation numbers;
 $c_r=1$ indicates that at site $r$ a spin
is present, whereas $c_r=0$ when this site is empty.
We shall consider uncorrelated
disorder with a fraction $c$ and a total number $cN$
of spins present, with $0<c<1$. $J_{rr'}=J(r-r')$
is a translationally invariant pair coupling, a well known example
being  the RKKY coupling of metallic spin glasses,
$J(r)\sim \cos(2k_Fr)/r^3$. However, we can consider more
general situations. In a metallic SG there may be additional
short-distance (anti-) ferromagnetic clustering \cite{Mydoshboek}.
For many insulating spin glasses $J(r)$ takes the value $J_{nn}>0$
for nearest neighbors and $J_{nnn}<0$ for next-nearest
neighbors (nnn), while it is effectively zero for more distant pairs.

In the following we shall focus on the situation where for
large $c$ ferromagnetic (FM) ordering occurs. For describing
two-sublattice antiferromagnetic ordering one first has to
make the
replacement $s_r\to -s_r$ on one of the sublattices.
This will redefine the couplings $J$
and introduce the staggered external field $\pm H$.
Since our spins are classical, this leads to a problem
similar to the ferromagnetic one.

In order to study the thermodynamics of this system at temperature
$T\equiv 1/\beta$ we
express the Boltzmann factor as a multiple integral
\BEQ
\exp\{-\beta {\cal H}(s)\}=\int D\phi \exp\{-\beta{\cal H}(\phi,s)\}
\EEQ
where
\BEQ
D\phi={\rm det}(\beta J/2\pi+i0)^{1/2}\prod_r \int d\phi_r
\EEQ
The integrations run from $-\sqrt{i}\infty$ to $\sqrt{i}\infty$.
To assure convergence of the Gaussian integrals,
a regulator $i0$ has been added to $J$; it will from now on not be
written explicitely.
The partition sum can thus be represented as trace over both discrete ($s_r$)
and continuous ($\phi_r$) degrees of freedom
\begin{equation}
Z=\int D\phi\sum_{\{s\}}
\exp\{-\beta{\cal H}(\phi,s)\},
\end{equation}
 with Hamiltonian
\begin{equation}
\beta {\cal H}(\phi,s)=\frac{T}{2}\sum_{rr'}\phi_r
J^{-1}_{rr'}\phi_{r'}-\sum_r c_r\{\phi_r+h\}s_r,
\end{equation}
where $h=\beta H$.

 The quenched averaged free energy follows from $\overline{\log Z}$.
The replica method is employed for studying this quantity.
One thus calculates the quenched averages $\overline{Z^n}$ for
 $n=1,2,3,\cdots$. The results have to be continued to the limit
 $n\to 0$. This procedure introduces replicated field and spin variables
 $\phi_r^\alpha$, $s_r^\alpha=s^\alpha_{r}$,
 for $1\le\alpha\le n$. The
average over the occupation variables $c_r$
can now be performed. Since spin sums are decoupled at each site,
we may now write  $s_r^\alpha$ as $s_\alpha$, which leads to
\begin{equation}\label{Znav}
\overline{Z^n}=\int D\phi \exp\{-\beta {\cal H}_n(\phi)\},
\end{equation}
 with replicated Hamiltonian
\begin{equation}\label{Hrepphi}
\beta {\cal H}_n(\phi)=\frac{T}{2}\sum_{\alpha=1}^n\sum_{rr'}\phi^\alpha_r
J^{-1}_{rr'}\phi^\alpha_{r'}-cN\Phi_n.
\end{equation}
where
\begin{equation}\label{Phn}
\Phi_n=n\log{2}+\frac{1}{cN}\sum_r
 \log\left(1-c+c\,{\rm tr}_s \exp{\sum_{\alpha=1}^n
(\phi^\alpha_r+h)s_\alpha}\right)\end{equation}
and where tr$_s$ stands for the normalized sum over $s_\alpha=\pm 1$, $(\alpha=1,
\cdots,\,n)$,
\BEQ
{\rm tr}_s=\frac{1}{2^n}\sum_{\{s_\alpha=\pm 1\}}=
\frac{1}{2^n}\sum_{s_1=\pm 1}\cdots \sum_{s_n=\pm 1}.
\EEQ

In order to proceed one has to make an assumption for the order
parameter(s). In a mean field approach the
 magnetization per spin, $M$, will be
proportional to the mean field value of $m_\alpha=
\left[\phi_\alpha\right]$, where
\begin{equation}
[A]\equiv N^{-1}\sum_r A_r
\end{equation}
denotes the spatial average of an observable $A_r$.
The main step in the present work is to introduce,
in the spirit of Edwards-Anderson, also space-independent composite order
parameters such as $[\phi_\alpha\phi_\beta]_c$.
This involves the second spatial cumulant
\BEQ
p_{\alpha\beta}=
[\phi_\alpha\phi_\beta]_c\equiv
\left[\phi_\alpha\phi_\beta\right]-
\left[\phi_\alpha\right]\left[\phi_\beta\right].
\EEQ
 $m_\alpha$ and  $p_{\alpha\beta}$ are the first terms in
 a cumulant expansion in powers of $\phi_\alpha$.
In Fourier space $m_\alpha$ equals $\hat\phi_\alpha(0)$
and $p_{\alpha\beta}=\sum_{k\neq 0}\hat\phi_\alpha(k)
\hat\phi_\beta(-k)$. Ferromagnetic
ordering implies a macroscopic occupation of the $k=0$ mode,
$\hat\phi_\alpha(0)={\cal O}(1)$, while a spin density wave
has a macroscopic occupation of some $k\neq 0$. Spin glass ordering,
 however, occurs when $p_{\alpha\beta}>0$ for
 $\alpha\neq \beta$ and is caused by a small
 occupation of all wavenumbers $k\neq 0$.

In our approach the order parameters explicitly take into account the
macroscopic contributions of combinations of Fourier modes.
(e.g. $M_\alpha=\hat\phi_\alpha(k=0)$).
The resulting Fourier sums (e.g.
$\sum_{k\neq 0}\hat\phi_\alpha(k)\hat\phi_\beta(k)$)
``cannot bite'', as their ``teeth''
have been removed. Likewise, at third order
\BEQ
[\phi_\alpha\phi_\beta\phi_\gamma]_c=
\sum_{k_1\neq 0;k_2\neq 0;k_1\neq-k_2
}\hat\phi_\alpha(k_1)\hat\phi_\beta(k_2)\hat\phi_\gamma(-k_1-k_2)
\EEQ
only incorporates small contributions from all $k_1$ and $k_1$, and
no  macroscopic ($M_\alpha$ or $q_\ab$) contribution.

For a full description of the problem an infinity of spatial cumulants is
needed. We shall, however, first see in how far low order approximations
already lead to meaningful results.

\subsubsection{Second order cumulant expansion or Gaussian approximation}
\label{2.1}

As in the derivation of the Fokker-Planck equation, one may hope that
the Gaussian approximation already leads to meaningful insights,
and does not yet produce unphysical effects such as negative entropies.

After expanding the logarithm of eq. (\ref{Phn}) in powers of
$c/(1-c)$, we introduce in the $\ell$'th term the
 concentration factor
\begin{equation}
\gamma_\ell=\frac{(-c)^{\ell-1}}{\ell(1-c)^\ell}\qquad
\qquad(\ell=1,2,3,\cdots)\end{equation}
 and we have to replicate the spins $\ell$ times. This
leads to the ``$\ell$-spin''
\begin{equation}\label{lspin}
\sigma_\alpha=\sum_{j=1}^\ell s_{\alpha}^{(j)}.
\end{equation}
The normalized trace over its values will be denoted
by ${\rm tr}_s^{(\ell)}$. When truncating the cumulant
expansion beyond second order, the spatial average leads to
\begin{equation}
\left[\exp\sum_\alpha\phi_\alpha\sigma_\alpha\right]\equiv
\exp[\exp\sum_\alpha\phi_\alpha\sigma_\alpha]_c=
\exp\left(\sum_\alpha m_\alpha\sigma_\alpha+\frac{1}{2}
\sum_{\alpha\beta}p_{\alpha\beta}\sigma_\alpha
\sigma_\beta\right).\end{equation}
The relation between
$p_{\alpha\beta}$ and the $\phi$'s is imposed by inserting
 for each set $\alpha\le\beta$
\BEA\label{1=pq}
1&=&\int \d p_{\alpha\beta}\delta(p_{\alpha\beta}-
[\phi_\alpha\phi_\beta]_c) \nn
&=&\int_{-\infty}^\infty \d p_{\alpha\beta}
\int_{-i\infty}^{i\infty}\frac{Nc}{4\pi i}\d q_\ab
\e^{\frac{1}{2}Ncq_\ab([\phi_\alpha\phi_\beta]_c-p_{\alpha\beta})}.
\EEA
A Gaussian form in the $\phi$'s is then obtained, so they can
be integrated out. This yields for the replicated free energy per spin
$\beta F_n=(-1/Nc)\ln Z_n$:
\begin{equation}
\label{Hmf}
\beta F_n
=\frac{T}{2c{\hat J}(0)}\sum_\alpha
m_\alpha^2 + \frac{1}{2c}\sum_\alpha
\int\dk\left\{\ln(1-c\beta\hat J(k)q)\right\}_{\alpha\alpha}+
\frac{1}{2}\sum_{\alpha\beta}
q_{\alpha\beta}p_{\alpha\beta}-\Phi_n,
\end{equation}
with
\begin{equation}\label{Phn=}
\Phi_n=n\log{2}+\sum_{\ell=1}^\infty \gamma_\ell
\left ({\rm tr}_s^{(\ell)}\exp{X^{(\ell)}}-1\right).
\end{equation}
Further,
\begin{equation}\label{Xl2oc=}
X^{(\ell)}=\sum_\alpha(m_\alpha+h)\sigma_\alpha+
\frac{1}{2}\sum_{\alpha\beta}
p_{\alpha\beta}\sigma_\alpha\sigma_\beta.
\end{equation}
The $k$-integral has arisen from $(1/N)\sum_{k\neq 0}$, where the exclusion
of the $k=0$ term is due to the definition of the spatial cumulant. In
the thermodynamic limit this exclusion becomes irrelevant.
The ensemble averaged free energy
per spin now follows as $F=\lim_{n\to 0}\, F_n/n$,
with $F_n$ calculated at its saddlepoint.
For $c\ge \frac{1}{2}$ the $\ell$-sum in eq. (\ref{Phn=}) can
defined by a Pad\'e resummation.

The magnetization per spin in the replicated system is
\begin{eqnarray}
M_\alpha&\equiv& \frac{T}{c\hat J(0)} m_\alpha=\overline{<\sigma_\alpha>}
\\
&\equiv& \sum_{\ell=1}^\infty\gamma_\ell\,\, {\rm tr}_s^{(\ell)}\,\,
\sigma_\alpha\,\exp{X^{(\ell)}}.
\end{eqnarray}

The physical interpretation of $q$ follows from its mean field
equation. This turns out to be very similar to the one
introduced by Edwards-Anderson and Sherrington-Kirkpatrick
\begin{equation}\label{Q=}
q_{\alpha\beta}=\overline{\langle \sigma_\alpha \sigma_\beta\rangle}
\equiv \sum_{\ell=1}^\infty \gamma_\ell
{\rm tr}_s^{(\ell)} \sigma_\alpha\sigma_\beta\exp{X^{(\ell)}}
=\sum_{\ell=1}^\infty \gamma_\ell
\frac{{\rm tr}_s^{(\ell)} \sigma_\alpha\sigma_\beta
\exp{X^{(\ell)}}}{{\rm tr}_s^{(\ell)} \exp{X^{(\ell)}}}.
\end{equation}
The last equality holds since the denominator
 equals unity in the replica limit $n\to 0$.
For $c\to 0$ only the $\ell=1$ term
$\overline{\langle s_\alpha s_\beta\rangle}$
survives and one exactly recovers the two state overlap introduced
by Edwards-Anderson and Sherrington-Kirkpatrick.
 Variation with respect to $q_{\alpha\beta}$ yields
\BEQ\label{p=kint}
p_{\alpha\beta}=\int\frac{{\rm d}^dk}{(2\pi)^d}\left(
\frac{\beta\hat J(k)}{1-c\beta \hat J(k) q}\right )_{\alpha\beta}.
\EEQ
This form  exhibits clustering effects.
Indeed, decomposing $q_{\alpha\beta}=q_d\delta_\ab+\tilde q_\ab$ with
$q_d=q_{\alpha\alpha}$ and $\tilde q_{\alpha\alpha}=0$,
we may also write for $\alpha\neq\beta$
\BEA
p_{\alpha \beta} & = &
\int\frac{{\rm d}^dk}{(2\pi)^d}\,\,
\beta\hat J_{\rm eff}(k)\left(\frac{1}{1-c\beta \hat J_{\rm
eff}(k) q}
\right )_{\alpha\beta}  \nn
 & = &
\int\frac{{\rm d}^dk}{(2\pi)^d}\,\,
c\beta^2\hat J^2_{\rm eff}(k)\left(\frac{\tilde q}{1-c\beta \hat J_{\rm
eff}(k)  q}
\right )_{\alpha\beta} \qquad{(\alpha\neq\beta)},
\label{p=kint2}
\EEA
with  the {\it effective coupling}
\BEQ \label{Jeff}
\hat J_{\rm eff}(k)=\frac{\hat J(k)}
{1-c\beta \hat J(k)q_d}.
\EEQ
Expansion in powers of $c$ indeed shows that it is a multi-spin
effect. Maxima of $\hat J$ are more pronounced in
$\hat J_{\rm eff}$. This describes the formation of finite
clusters that are precursors of the would-be spin density wave or
(anti-) ferromagnetic phase transition.
In metallic spin glasses incomplete spin density waves
have indeed been observed by neutron scattering.~\cite{Werner}
It is well known that
both in metallic and in insulating spin glasses there appear
ferromagnetic clusters near the ferromagnetic transition line; these
clusters (sometimes called ``fat spins'') act as a quite rigid effective
spins, that may contain up to 2000 magnetic atoms, making quite visible
moves. They  are responsible for the dynamics in the cluster glass phase
\cite{Mydoshboek}. We shall return to this point in Section~\ref{GLstuk}.

(At low temperature this expression for $J_{\rm eff}$
 becomes singular, though eq. (\ref{p=kint}) is well behaved.
It is then more appropriate to interpret
$\hat J/(1-c\beta\hat J(q_d-q_{EA}))$ as the effective coupling,
where $q_{EA}={\rm max}\,\, q_{\alpha\neq\beta}$
is the Edwards-Anderson order parameter).

\subsubsection{Replica symmetry}
To obtain an idea  of the content of previous expressions, we consider them
in the replica-symmetric sector. The free energy then reads
\BEA
\beta F&=&\frac{T}{2c\hat J(0)}m^2+\frac{1}{2c}\int\dk\left\{
\ln(1-c\beta\hat J(k)(q_d-q))
-\frac{c\beta\hat J(k)}{1-c\beta\hat J(k)(q_d-q)}
\right\}\nn
&+&\frac{1}{2}(p_dq_d-pq)-\sum_\ell\gamma_\ell \int g(x)\d x
\ln\int g(y)\d y 2^\ell \cosh^\ell(h+m+x\sqrt{p}+y\sqrt{p_d-p}),
\EEA
where
\BEQ g(x)=\frac{e^{-x^2/2}}{\sqrt{2\pi}}
\EEQ
is the Gaussian weight.
The saddle point equations read
\BEA
p&=&\int\dk \frac{c\beta^2 \hat J^2(k)}{(1-c\beta\hat J(k)(q_d-q))^2} ,\nn
p_d&=&p+\int\dk \frac{\beta \hat J(k)}{1-c\beta\hat J(k)(q_d-q)} , \nn
q&=&\sum_\ell\gamma_\ell \int g(x)\d x \ell^2
\left( \frac{\int g(y)\d y \cosh^\ell\psi
\tanh\psi}
{\int g(y)\d y \cosh^\ell\psi}\right)^2 ,\nn
q_d&=&\sum_\ell\gamma_\ell \int g(x)\d x
 \frac{\int g(y)\d y \cosh^\ell\psi
(\ell+\ell(\ell-1)\tanh^2\psi)}
{\int g(y)\d y \cosh^\ell\psi},
\EEA
where
\BEA
\psi & = & h+m+x \sqrt{\pi}+y \sqrt{p_d-p}.
\EEA
These expressions will be used later on for small $T$, where the
$\cosh$'s essentially become exponentials.

\subsubsection{Relation to the Hopfield model and the
Sherrington-Kirkpatrick model}

Despite of the simplifications made, the above expressions
are very rich. In  some limits of long range couplings
they  become exact.

Consider the situation where
\BEQ
J(k)=J_0 \qquad k_0<|k|<k_1
\EEQ
and zero outside this shell. This corresponds to a long range
spatial coupling
\BEQ
J(r)=J_0\int_{k_0<k<k_1} \frac{{\rm d}^dk}{(2\pi)^d} \frac{\sin(kr)}{kr}.
\EEQ
We denote the volume in phase space of this
shell by $\Delta=\int_{k_0}^{k_1}{\rm d}^dk/(2\pi )^d$. Consider the
combined limit $c\to 0$, $k_1\to k_0$, such that
$  \alpha=\Delta/c$ remains fixed.
Inserting this in eq. (\ref{Hmf})  we obtain for its spin glass content
($m=0$)
\begin{equation}
\label{Hopf1}
\beta F_n=\frac{\alpha}{2}\sum_\alpha
\left\{\log(1-c\beta J_0q)\right\}_{\alpha\alpha}+
\frac{1}{2}\sum_{\alpha\beta}
q_{\alpha\beta}p_{\alpha\beta}-\Phi_n.
\end{equation}
In order to understand its meaning, let us consider the case of
replica symmetry. We get in the limit $n\to 0$
\BEA
\label{Hopf2}
\beta F & = &  \frac{  \alpha}{2}\left\{
\log(1-c\beta J_0(1-q))-\frac{c\beta J_0q}{1-c\beta J_0(1-q)}\right\}
+\frac{1}{2}p(1-q) \nn
 & & -\int\frac{dx}{\sqrt{2\pi}}
e^{-x^2/2}\log{2\cosh x\sqrt{p}},
\EEA
with $p$ and $q$ obeying
\BEA\label{peqn}
 p&=&\frac{ \alpha (c\beta J_0)^2q}{(1-c\beta J_0(1-q))^2}, \nonumber\\
q&=&\int_{-\infty}^\infty
\frac{ {\rm d}x}{\sqrt{2\pi}}\,\,e^{-x^2/2}\,\tanh^2 x\sqrt{p}.
\EEA
After a rescaling ($T\to cJ_0\tilde T$, $p\to\alpha r/\tilde T^2$)
these are  exactly the replica-symmetric free energy and its saddle point
equations for the spin glass phase in the Hopfield model of a
neural network with $ \alpha N$ patterns ~\cite{Hopfield}.
 More generally, eq. (\ref{Hopf1}) is
the free energy including replica symmetry breaking. In the limit
$\Delta\to 0$, $c\to 0$, with fixed $\Delta/c=  \alpha$ one
considers long range interactions between widely separated spins,
that effectively lead to random interactions. It has been checked that
loop-corrections do not contribute in the limit. So (\ref{Hopf1}) is
an exact mean field theory in this long range limit.
Note that this may occur in any fixed dimension ($d=1,2,3,\cdots$).

The SK-model arises in the subsequent limit $ \alpha\to\infty$.
To show this we must rescale $T\to \tilde T cJ_0\sqrt{ \alpha}$.
Then the first relation in eq. (\ref{peqn}) becomes linear,
and eq. (\ref{Hopf1}) and
(\ref{Hopf2}) become linear+quadratic in $q_{\alpha\beta}$, so that
$p_{\alpha\beta}$ and $q_{\alpha\beta}$ are linearly related.
After elimination the $p$'s, we recover
expressions that in the limit $n\to 0$ are equivalent to the SK model.
Hereto we also use that for small $n$ our expression
$-1+{\rm tr}_s\exp(p_{\alpha\beta} s_\alpha s_\beta/2)$
is proportional to $n$ and to this order equivalent to the SK result
$\log{\rm tr}_s\exp(p_{\alpha\beta} s_\alpha s_\beta/2)$.

The present random site problem was also studied, for the
case of Heisenberg spins, by Bray and Moore \cite{BMrs}.
These authors do not introduce conjugated spin variables but
make a cumulant expansion of the Boltzmann factor up to second
order in the occupation factors $c_i$.
Since the latter are zero or one, there is, however,
no reason to stop at second order.
Therefore the results differ from ours and are incorrect.
For instance, according to ref. \cite{BMrs} the spin glass transition
 temperature scales as $T_g\sim c$ for small $c$ both for short
 range and RKKY couplings. We find, however, from eq. (\ref{peqn}) that
 $T_g=J_R\sim \sqrt{c}$. Below we shall derive $T_g\sim c$ for
RKKY couplings via a more intricate analysis, that takes into
account more contributions.

\subsubsection{Full cumulant expansion: the direct way}

The above results do not exhibit a percolation transition.
Indeed,  $T_{sg}\sim\sqrt{c}$ does not vanish
identically for small enough concentrations. This is related to the
fact that the
percolation transition of systems with $z=\infty$
occurs at $c=0$. Also for RKKY couplings the
present prediction overestimates $T_g$. Here one
expects the well established ``concentration scaling''
$T_{sg}\sim c$.\cite{Mydoshboek}

We therefore consider the full cumulant expansion
of eq. (\ref{Phn}). We introduce order parameters
\begin{equation}
p_{\alpha_1\cdots\alpha_k}=[\phi_{\alpha_1}\cdots\phi_{\alpha_k}]_c
\end{equation}
as well as conjugated parameters $q_{\alpha_1\cdots\alpha_k}$.
They are symmetric in their replica indices.
The replicated free energy may be written as
\begin{equation}\label{bFpq}
\beta F_n=-\Psi_n(q)+pq-\Phi_n(p),
\end{equation}
where
\begin{eqnarray}\label{Psi=}
\Psi_n=\frac{1}{cN}\log\int D\phi&\exp&\left(-\frac{T}{2}
\sum_{\alpha rr'}\phi^\alpha_rJ^{-1}_{rr'}\phi^\alpha_{r'}+
cN \sum_{k=1}^\infty\frac{1}{k!}
\sum_{\alpha_1\cdots\alpha_k}
q_{\alpha_1\cdots\alpha_k}[\phi_{\alpha_1}\cdots \phi_{\alpha_k}]_c\right),
\end{eqnarray}
and
\begin{equation}
pq= \sum_{k=1}^\infty\frac{1}{k!}\sum_{\alpha_1\cdots\alpha_k}
p_{\alpha_1\cdots\alpha_k}q_{\alpha_1\cdots\alpha_k}.
\end{equation}
$\Phi_n$ is given by eq. (\ref{Phn=}) with
\begin{equation}\label{X=}
X^{(\ell)}=\sum_\alpha (p_\alpha+h)\sigma_\alpha
+\frac{1}{2!}\sum_{\alpha\beta}p_{\alpha\beta}\sigma_\alpha\sigma_\beta+
\frac{1}{3!}\sum_{\alpha\beta\gamma}
p_{\alpha\beta\gamma}\sigma_\alpha\sigma_\beta\sigma_\gamma+\cdots.
\end{equation}
The free energy is now determined by the infinite set of
coupled equations
\begin{eqnarray}
p_{\alpha_1\cdots\alpha_k}&=&k!\frac{\partial}
{\partial q_{\alpha_1\cdots\alpha_k}}\Psi_n(q) \nonumber\\
q_{\alpha_1\cdots\alpha_k}&=&k!\frac{\partial}
{\partial p_{\alpha_1\cdots\alpha_k}}\Phi_n(p).
\end{eqnarray}

\subsubsection{A resummation}
\label{pViBr}
The above result is not very useful at low temperatures, where the
$p$'s become very large. However, it is possible to resum the contributions.
This can be performed in the following elegant way.
In the $\ell$'th
term of $\Phi_n$ there are spin variables $s_\alpha^{(j)}$,
$(j=1,\cdots, \ell)$, which we now label as $s_a$ with
$a=(\alpha,j)$. Though they do not
depend on $j$, we also denote $\phi_\a$ as $\phi_a$.
We start from the  definition
\begin{equation}\label{aj}
[\exp\sum_a\phi_\alpha s_a]=\exp\left([\exp\sum_a\phi_\alpha s_a]_c .
-1\right)\end{equation}
The obvious relation
\BEQ\label{lincum}
[A+A_a\phi_r^a+A_{ab}\phi_r^a\phi_r^b+A_{abc}\phi_r^a\phi_r^b\phi_r^c
+\cdots ]_c= A+A_a[\phi^a]_c+A_{ab}[\phi^a\phi^b]_c+
A_{abc}[\phi^a\phi^b\phi^c]_c+\cdots
\EEQ
(summation convention employed)
defines cumulants of any expandable local function of $\phi_r^a$.
It is a generalization of the definition
\BEQ\label{cumdef}
[\e^{\phi_as_a}]_c=1+s_a[\phi_a]_c +\frac{1}{2}s_as_b[\phi_a\phi_b]_c
+\frac{1}{6}s_as_bs_c[\phi_a\phi_b\phi_c]_c
+\frac{1}{24}s_as_bs_cs_d[\phi_a\phi_b\phi_c\phi_d]_c
+\cdots .
\EEQ
We employ the identity
\begin{equation}\label{cata}
\exp(\sum_a\phi_a s_a)
=\prod_a\cosh{\phi_a}(1+s_a\tanh{\phi_a})
\equiv \prod_a c_a(1+s_a t_a).
\end{equation}
Using the linear property of series of
cumulants, eq. (\ref{lincum}), this product brings
\BEA
[\exp\sum_a\phi_a s_a]_c
&=&[\prod_e c_e]_c
+\sum_a s_a[ t_a\prod_ec_e]_c
+\frac{1}{2}\sum_{a\neq b}s_as_b [ t_a t_b\prod_ec_e]_c
\nn
&+&\frac{1}{6}\sum_{a\neq b\neq c\neq a}
s_as_bs_c[t_a t_b t_c\prod_e c_e]_c
+\frac{1}{24}\sum_{a,b,c,d;{\em different}}
s_as_bs_cs_d[t_a t_b t_c t_d\prod_e c_e]_c
+\cdots .
\EEA
That the subtracted terms indeed work out this way can be checked
for low orders in the small $\phi_\a$ expansion. Up to quartic order
the right hand side gives
\BEA
&[1&+\frac{1}{2}\phi_a^2+\frac{1}{8}\phi_a^2\phi_b^2-\frac{1}{12}\phi_a^4]_c
+[s_a\phi_a(1-\frac{1}{3}\phi_a^2+\frac{1}{2}\phi_b^2)]_c
\nn
&+&\frac{1}{2}[s_as_b\phi_a\phi_b(1-\frac{1}{3}\phi_a^2-\frac{1}{3}\phi_b^2
+\frac{1}{2}\phi_c^2)-\phi_a^2(1-\frac{2}{3}\phi_a^2+\frac{1}{2}\phi_c^2)]_c
\nn
&+&\frac{1}{6}[s_as_bs_c\phi_a\phi_b\phi_c-3s_a\phi_a\phi_b^2+2s_a\phi_a^3]_c
\nn
&+&\frac{1}{24}[s_as_bs_cs_d\phi_a\phi_b\phi_c\phi_d-6s_as_b\phi_a\phi_b\phi_c^2
+3\phi_a^2\phi_b^2+8s_as_b\phi_a\phi_b^3-6\phi_a^4]_c , 
\EEA
where now the sums over repeated indices are unrestricted.
This expression indeed coincides
 with eq. (\ref{cumdef}).

 For every integer  $1\le k \le n$ and $\ell\ge 1$ we introduce symmetric
order parameters
\begin{equation}
p^{(\ell)}_{\alpha_1\cdots\alpha_k}
=[c_1^\ell\cdots c_n^\ell
t_{\alpha_1}\cdots t_{\alpha_k}]_c,
\end{equation}
with the restriction that the total number of replica indices that occur
more than once, does not exceed $\ell$. (For example, $p^{(2)}_{\alpha\alpha
\beta\gamma}$  is needed, but $p^{(4)}_{\alpha\alpha\alpha
\beta\beta\gamma}$ is not). Further
\BEQ
p^{(\ell)}=[c_1^\ell\cdots c_n^\ell]_c-1 . \EEQ
Introducing the spin variable
\begin{equation}\label{Xl=}
X^{(\ell)}=\sum_a (p_\alpha^{(\ell)}+h)s_a
+\frac{1}{2!}\sum_{a\neq b}p^{(\ell)}_{\alpha\beta}s_a s_b+
\frac{1}{3!}\sum_{a\neq b\neq c\neq a}
p^{(\ell)}_{\alpha\beta\gamma}s_a s_b s_c+\cdots,
\end{equation}
we find that the $\ell$'th term of $\Phi_n$ involves an expression
${\rm tr}_s^{(\ell)}\exp\{p^{(\ell)}+X^{(\ell)}\}$.
Next we introduce the conjugate order parameters
$q^{(\ell)}$, $ q^{(\ell)}_\alpha$, $q^{(\ell)}_\ab$, $\cdots$,
through relations of the type (\ref{1=pq}),
\BEA\label{1=pqkl}
1&=&\int_{-\infty}^\infty \d p^{(\ell)}
_{\alpha_1\cdots\alpha_k}
\int_{-i\infty}^{i\infty}\frac{Nc}{k!2\pi i}
\d q^{(\ell)}_{\alpha_1\cdots\alpha_k}
\exp{\frac{1}{k!}Nc\,q^{(\ell)}_{\alpha_1\cdots\alpha_k}
([t_{\alpha_1}\cdots t_{\alpha_k}\prod_ec_e]_c
-p^{(\ell)}_{\alpha_1\cdots\alpha_k})}
\EEA

 Let us first consider
the saddle point equations for $p^{(\ell)}$ and $q^{(\ell)}$.
It is readily seen that $ p^{(\ell)}={\cal O}(n)$ and that
$q^{(\ell)}=1+{\cal O}(n)$. This implies that $p^{(\ell)}$
can be omitted and that the $q^{(\ell)}$ terms can be summed.
These manipulations lead us to
a replicated free energy of the form (\ref{bFpq}),
\begin{equation}\label{bFVBc}
\beta F_n=-\Psi_n(q)+(p,q)-\Phi_n(p),
\end{equation}
with $\Phi_n$ given by eqs. (\ref{Phn=}) and (\ref{Xl=}). Further,
\begin{equation} \label {pqVB}
(p,q)\equiv \sum_{\ell=1}^\infty\gamma_\ell\sum_{k=1}^\infty \frac{1}{k!}
\sum_{\alpha_1\cdots\alpha_k}p^{(\ell)}_
{\alpha_1\cdots\alpha_k}
q^{(\ell)}_{\alpha_1\cdots\alpha_k}
\end{equation}
and
\begin{eqnarray}\label{Psi2=}
\Psi_n=\frac{1}{cN}\log\int D\phi&\,&\exp\left(-\frac{T}{2}
\sum_{\alpha rr'}\phi^\alpha_rJ^{-1}_{rr'}\phi^\alpha_{r'}
+N\left[\log\left(1-c+c\,{\rm tr}_s
\exp\sum_\alpha \phi_\alpha s_\alpha\right)
\right]_c\right) \nonumber\\
\ast&&\exp\left(cN\sum_{\ell=1}^\infty\gamma_\ell
\sum_{k=1}^\infty\frac{1}{k!}\sum_{\alpha_1\cdots\alpha_k}
q^{(\ell)}_{\alpha_1\cdots\alpha_k}[c_1^\ell\cdots c_n^\ell
t_{\alpha_1}\cdots t_{\alpha_k}]_c\right).
\end{eqnarray}
In this expression the logarithm sums the $q^{(\ell)}\to 1$ contributions.
The difficult part of the problem is still  the integration
over the $\phi$ fields. One might even think that nothing has been
gained in the present representation. However, the dangerous order
parameters have been subtracted explicitely. In the remainder
of the paper we shall mainly consider situations where $\Psi_n$ has
 a relatively simple form. Nevertheless, we shall be able
to draw interesting conclusions.

In general, the $\phi$ integrals cannot be
performed exactly. If the exponent is expanded up to
 second order in $\phi$, one recovers eq. (\ref{Hmf}).

\subsubsection{Small concentrations and the
relation with the Viana-Bray model}
\label{ViBr}

We now consider the limit of small $c$ at fixed $T$. Here only $\ell=1$
contributes, so that all
 $q$'s and $p$'s with $\ell>1$ can be omitted.
The remaining ones have all replica
indices different from each other. $\Psi_n$ can be approximated
to order $q^2$. The integrals can be performed. To see how this goes,
consider eq. (\ref{Psi2=}). First, since $c$ is small, we may
omit the term $N[\log\cdots]_c$ in the exponent.
When expanding in powers of $q$ we have bilinear terms of the type
\begin{eqnarray}\label{Psi2ontw}c^2q_{\alpha_1\cdots\alpha_k}
q_{\alpha_1'\cdots\alpha_{k'}'}
\int D\phi\exp(-\frac{T}{2}
\sum_{\alpha rr'}\phi^\alpha_rJ^{-1}_{rr'}\phi^\alpha_{r'})
N[c_1\cdots c_n t_{\alpha_1}\cdots t_{\alpha_k}]_c
N[c_1\cdots c_n  t_{\alpha_1'}\cdots t_{\alpha_{k'}'}]_c .
\end{eqnarray}
Neglecting, for the moment, the difference between cumulants and
ordinary moments, the integral can be written as
\BEA &\,&
\sum_{r_1,r_2}\sum_{s,\tilde s}
s_{\alpha_1}\cdots s_{\alpha_k}
\tilde s_{\alpha_1'}\cdots \tilde s_{\alpha_{k'}'}
\int D\phi\exp(-\frac{T}{2}
\sum_{\alpha rr'}\phi^\alpha_rJ^{-1}_{rr'}\phi^\alpha_{r'})
\exp\sum_\gamma( \phi_{r_1}^\gamma s _\gamma+
\phi_{r_2}^\gamma \tilde s_\gamma)\nonumber\\
&=&\sum_{r_1,r_2}\sum_{s,\tilde s}
s_{\alpha_1}\cdots s_{\alpha_k}
\tilde s_{\alpha_1'}\cdots \tilde s_{\alpha_{k'}'}
\exp(\beta J(r_1-r_2)\sum_\gamma s_\gamma\tilde s_\gamma) .
\EEA
For each fixed $\gamma$,  the sum over $s_\gamma$ and $\tilde s_\gamma$
can have preexponential factor $1$, $s_\gamma$,
$\tilde s_\gamma$, and $s_\gamma\tilde s_\gamma$. When performing the sums,
it is readily see that only the first or the last type of terms survive;
this explains also why terms linear in $q$ vanish identically.
This implies that only diagonal terms in the
$q_{\alpha_1\cdots\alpha_k}$ remain, with prefactors that are
spatial sums of $\cosh^n\beta J(r_1-r_2)\tanh^k\beta J(r_1-r_2)$.
Investigating the subtraction terms that occur in the spatial cumulants
we observe that replacing the cumulants by ordinary moments leads
at this point only to corrections of order $1/N$, that are negligible
in the thermodynamic limit. The same effect led to
the exclusion of the $k=0$ term in (\ref{Hmf}).
 This supports previous finding that
the role of the order parameters is only to eliminate the macroscopic parts
of spatial averages; the remainder being the well-behaved cumulants.
For small $n$ we find for the replicated free energy
\begin{equation}\label{bFVB}
\beta F_n= \frac{c}{2}\left\{-n\tau_0+
\tau_1\sum_\alpha M_\alpha^2+
\tau_2\sum_{\alpha<\beta}q_{\alpha\beta}^2
+\tau_3\sum_{\alpha<\beta<\gamma}
q_{\alpha\beta\gamma}^2+\cdots\right\}
-n\log{2}+1-{\rm tr}_s\exp{X} ,
\end{equation}
with
\BEQ\tau_0=\sum_r\log\cosh\beta J(r)
\qquad\tau_k=\sum_r \tanh^k\beta J(r) \qquad( k\ge 1)
\EEQ
and
\begin{equation}\label{X2=}
X=\sum_\alpha(c\tau_1 M_\alpha+h) s_\alpha+
c\tau_2\sum_{\alpha<\beta}q_{\alpha\beta}s_\alpha s_\beta+c\tau_3
\sum_{\alpha<\beta<\gamma} q_{\alpha\beta\gamma}
s_\alpha s_\beta s_\gamma+\cdots .
\end{equation}
When couplings are (mainly) ferromagnetic,
a ferromagnetic transition sets in at temperature such that $c\tau_1=1$.
If the $J(r)$ are partly positive and negative, there are cancellations
in this sum. Then a spinglass phase may occur at a higher temperature.
The spinglass phase ($q_{\alpha\beta}>0$)
sets in at a temperature $T_g$ where $c\tau_2=1$.
If only couplings to $z$ neighbors are different from zero,
a SG phase can only occur beyond the percolation threshold $c_p=1/z$.
$c/c_p>1$ expresses that, on the average, each spin should
interact with more than one other spin.
For a fcc lattice in $d=3$ this yields
 $c_p=1/12\approx 8.3\%$, to be compared with the
simulated value $c_p\approx 13.6\%$ \cite{Essam}.
For the RKKY-coupling in three dimensions, the condition
 $c\tau_2=1$ indeed yields the ``concentration scaling'' $T_{G}\sim c$
for small  $c$.

Some years ago the SK model with diluted random bonds was introduced
by Viana and Bray\cite{VB}; it is commonly called the
Viana-Bray model. These authors also analyzed various aspects of the phase
diagram. In an unpublished work one of the present authors~\cite{Nunp}
derived the  same model, without making any analysis of the phase diagram.
In the diluted model one assumes that the couplings $J_{rr'}$ in
eq. (\ref{Hising}) vanish with probability $1-p/N$, and are
drawn independently from a normalized distribution $r(J)$ with
probability $p/N$. In the thermodynamic limit $N\to\infty$,
the parameter $p$ is kept fixed. There is a close analogy
with the Flory model of coagulation\cite{Flory}\cite{Hendriks}.
The parameter $p$ can be interpreted as a dimensionless time
variable, at which the coagulation has been interrupted. For $p<1$
only finite clusters exist, while for $p>1$ an infinite cluster has
appeared. On this infinite cluster a spontaneous
phase transition can occur, to a (anti-) ferromagnetic or spinglass
phase.

It would be interesting to consider $p$ really as a slow time
variable, and to consider the combination of very slow cluster growth
and spin glass dynamics. In this dynamical
model the random couplings would
be fixed, but as time progresses more and more of them would become
active. This idea is inspired on a model, introduced
by Coolen, Penney, and Sherrington, where all couplings are active,
but their strength changes slowly in time~\cite{Penney}.

The quenched free energy of the Viana-Bray model
is derived along the following lines.
The averaged replicated partition sum equals
\begin{eqnarray}
\overline{Z^n}&&=\sum_{\{s_r^\alpha\}}\prod_{<rr'>}\left\{
1-\frac{p}{N}+\frac{p}{N}\int r(J)dJ\exp(\beta J\sum_\alpha s^\alpha_r
s^\alpha_{r'})\right\}\nonumber\\
&&\approx\sum_{\{s_r^\alpha\}}\exp\left\{
\frac{p}{2N}\sum_{rr'}\left(-1+\int r(J)dJ
\cosh^n(\beta J)\prod_\alpha(1+ s^\alpha_r
s^\alpha_{r'}\tanh\beta J)
\right)\right\}.
\end{eqnarray}
Writing out the product over $\alpha$ and decoupling the
terms quadratic in products of spins at the same site, one defines
\begin{equation}
t_0=\int r(J)dJ\log\cosh\beta J;\qquad\qquad t_k=\int r(J)dJ\tanh^k\beta J
\qquad(k=1,2,3,\cdots).
\end{equation}
In the limit $n\to 0$ the replicated free energy takes the form
\begin{equation}\label{bFVBNu}
\beta F_n= \frac{p}{2}\left\{-nt_0+
t_1\sum_\alpha M_\alpha^2+
t_2\sum_{\alpha<\beta}q_{\alpha\beta}^2
+t_3\sum_{\alpha<\beta<\gamma}
q_{\alpha\beta\gamma}^2+\cdots\right\}
-n\log{2}-\log{\rm tr}_s\exp{X},
\end{equation}
with $X$ given by eq. (\ref{X2=}) after the replacement $c\tau_k\to pt_k$.
For a discussion of the phase diagram of this model, see e.g.
refs. \cite{VB},\cite{MdD},
\cite{dDG},\cite{AdDM}.

For short range couplings in the random site problem
the two results are thus closely related.
Indeed, since again $-1+{\rm tr}_s\exp X$ is equivalent
to $\log{\rm tr}_s\exp X$,
the only real difference lies in the meaning
of the parameters $c\tau_k=(c/c_p)(\tau_k/z)=pt_k$.
In the diluted random bond model $\tau_k/z=t_k$ is defined as
the average of $\tanh^k{\beta J}$ over the distribution of the
independent couplings $J$. In the random site model with sure
 couplings, this average is taken over space.

\section{Gaussian variational approach}

Since disordered systems are so hard to solve, it is becoming more
and more popular to investigate variational approaches
{}~\cite{HoneyThir}, ~\cite{MP},~\cite{OG},~\cite{BO}.
In the present problem one starts from eqs. (\ref{Hrepphi})
and (\ref{Phn}),
and chooses the variational Hamiltonian
\BEQ
{\cal H}_{\rm var}=\frac{1}{2}\sum_{ij\alpha\beta}(\phi_i^\alpha-m_\alpha)
\{G^{-1}\}_{ij}^{\alpha\beta}(\phi_j^\beta-m_\beta)
\EEQ
The variational propagator
$\langle\phi_i^\alpha\phi_j^\beta\rangle_{\rm var}
=G_{ij}^{\alpha\beta}
=G_\ab(r_i-r_j)$ is translationally invariant
and has Fourier transform $\hat G_\ab(k)$.
The variational free energy per spin becomes
\BEA\label{varfreeen}
\beta F&=&\frac{1}{Nc}\left(-\ln Z_{\rm var}
+\langle {\cal H}_n-{\cal H}_{\rm var}\rangle_{\rm var}\right)
\nonumber \\
&=&\sum_\alpha \frac{m_\alpha^2}{2\beta
\hat J(k=0)}+\frac{1}{2c}\sum_{\alpha}\int \dk \{-\ln \hat G(k)
+ \frac{\hat G(k)}
{\beta \hat J(k)}\}_{\alpha\alpha}
\nonumber\\
&+&\sum_{\ell=1}^\infty \gamma_\ell (1-\tr_{\sigma}^{(\ell)}
e^{\sum_\alpha (h+m_\alpha)\sigma_\alpha+\frac{1}{2}\sum_{\alpha\beta}
\sigma_\alpha G_{\alpha\beta}(r=0)\sigma_\beta}) .
\EEA
The variational equation $\delta F/\delta G_{\alpha\beta}(k)=0$ can be
expressed in terms of $q_\ab$ using (\ref{Q=}). This  yields
\BEQ
G_{\alpha\beta}(k)=\left(\frac{\beta \hat J(k)}
{1-c\beta \hat J(k)q}\right)_{\ab}.
\EEQ
Its $k$-integral can be identified with eq. (\ref{p=kint}):
\BEQ
p_\ab=G_\ab(r=0)=\int \dk  \left(\frac{\beta \hat J(k)}
{1-c\beta \hat J(k)q}\right)_{\ab} .
\EEQ
Due to this, the $\ell$-sum in eq. (\ref{varfreeen}) can be identified
with the expressions (\ref{Phn=}), (\ref{Xl2oc=}). Therefore the
variational equations of the second order cumulant
expansion~\cite{NEPL}
 and the Gaussian variational approximation coincide. It is then a
small exercise to check that the saddle point value of the free
energy also coincides. This does not mean that both approaches
are the same. The Gaussian variational Ansatz has looked in a
larger space, with variational parameters $G_\ab(r)$, but found
the same physics as the second order cumulant expansion
with its space-independent parameters $p_\ab$ and $q_\ab$.
The reason hereto is simply that both approximations are Gaussian.
The equivalence of the two approaches
 was realized by one of us in fall  1995.

One can go to the continuum limit. Hereto one reinserts
the lattice constant $a$, and takes the combined limit
$c\to 0$, $a\to 0$, such that $\rho=c/a^d$ is fixed. The main
effect is an overall factor $a^d$ (since in the limit $F/a^d$ becomes the
finite free energy per unit volume) and that only the $\ell=1$ term
survives in the sum (because $c\to 0$).
This continuum result was discovered independently
by Dean and Lancaster~\cite{DL}, who started in the continuum
immediately. (We should point out that a proper definition of
their path integral would reintroduce a lattice.)
They introduce a grand-canonical description of disorder, which is
physically the same as the ordinary disorder average,
but for their purpose a bit more convenient.

We have criticism to this work. The claim of ~\cite{DL} to present
a new field theoretic approach is unjust. Their field theory
in replica space (paper I, eq. (4), paper II eq. (2.5))
is just the continuum limit of the above eqs.
(\ref{Znav}), (\ref{Hrepphi}), (\ref{Phn}), represented
long ago as eqs. (2) and (3) of ref.~\cite{NEPL}.
The authors of \cite{DL} fully refrained from explaining
that, despite the different approach,  the
resulting  saddle point equations and saddle point free energy
are not new either ~\cite{DLRome}, but coincide
with the ones of the second order cumulant expansion
of \cite{NEPL}, and also discussed above.

\section{A site-disordered spin glass model with pair overlaps only}
\label{sitepaironly}
In realistic spin glasses the signs of the interactions
oscillate due to the strong distance-dependence for pairs of
spins present. We have seen that this leads to global multi-spin order
parameters, a somewhat unfamiliar theoretical framework.
On the other hand, it is custommary
to consider models where the signs are uncorrelated random variables.
We shall extend that approach to the site-diluted case, and find that only
local pair overlaps occur.

Consider the couplings
\BEQ
J_{rr'}=J(r-r')\xi_r\xi_{r'}
\EEQ
with independent quenched random numbers $\xi_r$.
Numerically the simplest choice is $\xi_r=\pm 1$ with equal probabilities.
We shall take them, however, Gaussian with zero average and unit variance.
The function $J(r)$ can now be taken non-negative.  For $J(r)\sim 1/r^3$
this form replaces the deterministic signs of the RKKY interaction by
random signs. We proceed  along the lines of section 2, and introduce
fields $\phi_r^\alpha$.
The disorder average over the random spin configurations and the
random $\xi$'s will now yield
\BEQ
\overline{\exp\left(\sum_\alpha s_\alpha c_r(\phi^\alpha_r\xi_r+h)\right)}
=1-c+c\,\,\exp\left(\frac{1}{2}\sum_{\alpha\beta}
\phi_r^{\alpha}\phi_r^\beta s_\alpha s_\beta +h\sum_\alpha s_\alpha\right)
\EEQ
We can introduce the composite local field
$p_{\alpha\beta}(r)=\phi_r^{\alpha}\phi_r^\beta$ as new variable
by inserting $\delta$ functions at each lattice site. In their plane
wave representations there appear only Gaussian $\phi$ integrals.
Integrating them out one is left with a field theory
for the space-dependent fields $p_{\alpha\beta}(r)$, $q_{\alpha\beta}(r)$.
Its Hamiltonian reads
\BEQ
\beta {\cal H}_n=\sum_r \beta {\cal H}(m(r),p(r),q(r))
\EEQ
with
\BEA \label{Hmf4}
\beta {\cal H}(m,p,q)&=&
\frac{T}{2c{\hat J}(0)}
\sum_\alpha m_\alpha^2 + \frac{1}{2c}\sum_\alpha
\{\log(1-c\beta\hat J(k)q)\}_{r\alpha;r\alpha}+
\frac{1}{2}c\sum_{\alpha\beta}q_{\alpha\beta}p_{\alpha\beta}
\nonumber\\ &-&
\ln\left(1-c+c{\rm tr}_s
\exp\left(\frac{1}{2}\sum_{\alpha\beta}
p_{\alpha\beta}
s_\alpha s_\beta +h\sum_\alpha s_\alpha\right) \right)
\EEA
Note that the log-term is non-local.
In the mean field approximation, the functions $p$ and $q$ are space-
independent. The resulting expression is then very close to the one in
the second order cumulant expansion of section 2.
Now the spin sums are simpler, since the logarithm
needs not be expanded. Consequently, the sign changes in the
coefficients $w$ and $y_1$, $y_2$, $y_3$, to be discussed in
section ~\ref{GLstuk}, do not occur in the present model.

As in section 2, eq. (\ref{Jeff}),
 the effective coupling is
\BEQ
\hat J_{\rm eff}(k)=\frac{\hat J(k)}{1-c\beta \hat J(k)(1-q_{EA})},
\EEQ
which again exhibits clustering effects.

\section{Vector spins}

Experimentally, spinglasses usually consist of Heisenberg spins with
weak or strong anisotropy. It is therefore useful to extend our
formalism to $m$-component spins. We consider the Hamiltonian
\begin{equation}
\label{HHeis}
{\cal H}(s)=-\frac{1}{2}
\sum_{\mu\nu rr'}J_{rr'}^{\mu\nu}S_r^\mu S_{r'}^\nu c_rc_{r'}-
\sum_{\mu r} H_r^\mu S_r^\mu c_r
\end{equation}
where $\vert {\bf S}\vert=1$.
The translationally invariant coupling
$J$ may contain effects of anisotropy, such as dipolar anisotropy.

\subsubsection{Second order cumulant expansion}
Proceeding as above we can carry out the calculations in the second
order cumulant expansion. We find
\begin{equation}
\beta F_n=\frac{\beta}{2c}\sum_{\alpha\mu\nu}\hat J^{\mu\nu}(0)
M_\alpha^\mu M_\beta^\nu+\frac{1}{2c}\int_k \sum_{\alpha\mu}
\{\log(1-c\beta\hat J(k)q)\}_{\alpha\mu;\alpha\mu}
+\frac{1}{2}\sum_{\alpha\beta\mu\nu}
p^{\mu\nu}_{\alpha\beta}q^{\mu\nu}_{\alpha\beta}-\Phi_n .
\end{equation}
The expression for $\Phi_n$ takes the form
\begin{equation}\label{Phinvector}
\Phi_n=n\log\Omega_m+\sum_{\ell=1}^\infty\gamma_\ell\left\{-1+
{\rm tr}_S^{(\ell)} \exp X^{(\ell)}\right\}.
\end{equation}
Here $\Omega_m$ is the area of a unit sphere in $m$ dimensions
($\Omega_1=2,\,\Omega_2=2\pi;\,\Omega_3=4\pi$).
$D{\bf S}=\Omega_m^{-n}d{\bf S}_1\cdots d{\bf S}_n$, is the
angular integration measure for one replicated spin and
${\rm tr}_S^{(\ell)}=\prod_{j=1}^\ell\int D{\bf S}^{(j)}$ is the
 integration over all replicated spins. Finally,
\begin{equation}
X^{(\ell)}=\sum_{\alpha\mu} (m_\alpha^\mu+h^\mu)\sum_{j=1}^\ell
S^{(j)\mu}_\alpha+\frac{1}{2}\sum_{\alpha\beta\mu\nu}
p_{\alpha\beta}^{\mu\nu}
\sum_{jj'=1}^\ell S_\alpha^{(j)\mu} S_\beta^{(j')\nu}.
\end{equation}

In the SK-type limit discussed above, one again obtains
and expression quadratic in the $q$'s:
\begin{equation}
\beta F_n=\frac{\beta}{2c}\sum_{\alpha\mu\nu r} J^{\mu\nu}_{r0}
M_\alpha^\mu M_\alpha^\nu+\frac{\beta^2}{4c}\sum_r
\sum_{\alpha\beta}\sum_{\mu\nu\rho\sigma}J_{0r}^{\sigma\mu}
J_{r0}^{\nu\rho}q_{\alpha\beta}^{\mu\nu}q_{\beta\alpha}^{\rho\sigma}
+\frac{1}{2}\sum_{\alpha\beta\mu\nu}
p^{\mu\nu}_{\alpha\beta}q^{\mu\nu}_{\alpha\beta}-\Phi_n .
\end{equation}
For isotropic couplings in the presence of an external field,
Gabay and Toulouse\cite{GT} already showed the occurrence of an
irreversibility line where transverse components of the spins freeze.

\subsubsection{The resummed cumulant expansion}

Here our aim is to derive the equivalent for vector spins of the
resummed cumulant expansion of section \ref{pViBr}.
This will yield, in particular,
a prediction for the transition temperature at low concentrations.

The first step in the derivation is the vector analog of the
cumulant generating function eq. (\ref{aj}):
\begin{equation}\label{aj3}
[\exp\sum_{a\mu}\phi_\alpha^\mu S_a^\mu]=
\exp\left([\exp\sum_{a\mu}\phi_\alpha^\mu S_a^\mu]_c
-1\right) ,\end{equation}
where $\mu=1,2,\cdots,m$ and where $a=(\alpha,j)$ with $\alpha=1,2,\cdots,n$
and $j=1,2,\cdots,\ell$.
The natural generalization of eq. (\ref{cata}) is to introduce
\begin{equation}
\exp{\sum_\mu\phi_\alpha^\mu S_a^\mu}=C_a+C_a
T_\alpha({\bf S}_a)=P_a\exp{\sum_\mu\phi_\alpha^\mu S_a^\mu}
+Q_a\exp{\sum_\mu\phi_\alpha^\mu S_a^\mu} ,
\end{equation}
where
\begin{equation}
C_a=\frac{1}{\Omega_m}\int d{\bf S}_a
\exp{\sum_\mu\phi_\alpha^\mu S_a^\mu}
\equiv P_a \exp{\sum_\mu\phi_\alpha^\mu S_a^\mu} ,
\end{equation}
with $P_a=\Omega_m^{-1}\int d{\bf S}_a$ being a projector,
$Q_a=1-P_a$ being its complement, and
\begin{equation}
T_\alpha({\bf S}_a)=\frac{\exp{\sum_\mu\phi_\alpha^\mu S_a^\mu}}
{C_\alpha}-1.
\end{equation}
In analogy with section \ref{pViBr} we define order parameter functions
\begin{equation}
p^{(\ell)}_{\alpha_1\cdots\alpha_k}({\bf S}_{a_1},\cdots,
{\bf S}_{a_k})=
\left[C_1^\ell\cdots C_n^\ell T_{\alpha_1}({\bf S}_{a_1})
\cdots T_{\alpha_k}({\bf S}_{a_k})\right] ,
\end{equation}
where the total number of repeated replica indices should not exceed $\ell$.
Note that this quantity depends only on $k$ spin variables, generalizing
a similar property of the combination
$p_{\alpha_1\cdots\alpha_k}s_{a_1}\cdots
s_{a_k}$ of section \ref{pViBr}.
The expression for $\Phi_n$ takes the form (\ref{Phinvector}) with
\begin{equation}
X^{(\ell)}=\sum_a \{p_\alpha^{(\ell)}({\bf S}_{a})+
{\bf h\cdot S}_a\}+\frac{1}{2!}\sum_{a\neq b}
p_{\alpha\beta}^{(\ell)}({\bf S}_{a},{\bf S}_b)+\frac{1}{3!}
\sum_{a\neq b\neq c\neq a}
p_{\alpha\beta\gamma}
^{(\ell)}({\bf S}_{a},{\bf S}_b,{\bf S}_c)
+\cdots.\end{equation}

Going through similar steps as before we derive the replicated free energy
\begin{equation}
\beta F_n=-\Psi_n(q) +pq-\Phi_n(p) ,
\end{equation}
where
\begin{eqnarray}\label{Psi2vs=}
\Psi_n=&&\frac{1}{cN}\log\int D\phi\exp\left(-\frac{T}{2}
\sum_{\alpha\mu r}\phi_\alpha^\mu(r)\{J^{-1}\phi_\alpha\}^\mu(r)
+N\left[\log\left(1-c+c\,\int D{\bf S}
\exp\sum_\alpha \phi_\alpha^\mu S_\alpha^\mu\right)
\right]_c\right) \nonumber\\
\ast&&\exp\left(cN\sum_{\ell=1}^\infty\gamma_\ell
\sum_{k=1}^\infty\frac{1}{k!}\sum_{\alpha_1\cdots\alpha_k}
P_1\cdots P_k\,
q_{\alpha_1\cdots\alpha_k}^{(\ell)}({\bf S}_1,
\cdots,{\bf S}_k)\left[C_1^\ell
\cdots C_n^\ell T_{\alpha_1}({\bf S}_1)
\cdots T_{\alpha_k}({\bf S}_k)\right]_c\right).
\end{eqnarray}
and where
\begin{equation}
pq= \sum_{\ell=1}^\infty\gamma_\ell\sum_{k=1}^\infty\frac{1}{k!}
\sum_{\alpha_1\cdots\alpha_k}
P_1\cdots P_k\,
p_{\alpha_1\cdots\alpha_k}^{(\ell)}({\bf S}_1,\cdots,{\bf S}_k)
q_{\alpha_1\cdots\alpha_k}^{(\ell)}({\bf S}_1,\cdots,{\bf S}_k) .
\end{equation}
In the limit of strong anisotropy, where the vector spins
are Ising-like and point
along the $\pm z$ axis, these equations reduce
to the results of section \ref{pViBr}. Indeed, it can be seen
that $q_{\alpha_1\cdots\alpha_k}^{(\ell)}({\bf S}_1,\cdots,{\bf S}_k)
\to \{q_{\alpha_1\cdots\alpha_k}^{(\ell)}\}_{\em Ising}
S_1^z\cdots S_k^z$ in this limit.

\subsubsection{The spin glass transition at low concentrations}
\label{TGMatIg}
The spin glass transition occurs for small
$p_{\alpha\beta}^{(\ell)}$ and $q_{\alpha\beta}^{(\ell)}$.
We now restrict ourselves to the case of small $c$, where only
the $\ell=1$ terms contribute. The transition temperature
is found by considering a replica symmetric solution
$q_{\alpha\beta}({\bf S}_1,{\bf S}_2)=q_{SG}({\bf S}_1,
{\bf S}_2)$. By expanding the mean field equations of
 $p_{\alpha\beta}$ and $q_{\alpha\beta}$
it is seen that the condition for the critical point has the form of
an integral equation
\begin{equation} \label{TGG}
 q_{SG}({\bf S}_1,{\bf S}_2)=c\sum_r \frac{
P_3 P_4\, q_{SG}({\bf S}_3,{\bf S}_4)Q_1Q_2Q_3Q_4
\exp\left(\sum_{\mu\nu}\{
S_1^\mu\beta \tilde J_{0r}^{\mu\nu}S_3^\nu+
S_2^\mu\beta \tilde J_{0r}^{\mu\nu}S_4^\nu\}\right) }
{\left\{ P_5P_6\exp\left(\sum_{\mu\nu}
S_5^\mu\beta \tilde J_{0r}^{\mu\nu}S_6^\nu\right)\right\}^2},
\end{equation}
where $\tilde J_{0r}^{\mu\nu}=\frac{1}{2}(J_{0r}^{\mu\nu}+J_{r0}^{\nu\mu})$.
Note that this equation has the form eigenvalue equation,
${\rm Kernel}\ast q_{SG}=\frac{1}{c}q_{SG}$.
This prediction is useful for long range spin glasses with a small
concentration of spins, as typically occurs in experiments on metallic
spinglasses and also in the Monte Carlo experiment of ref. \cite{MatIg}.

The fact that an integral equation occurs is not unexpected, since this
is already known to happen for a linear chain. Because of this reason,
the analysis of the SG phase for very diluted
vector spins is more cumbersome than for Ising spins.

For isotropic couplings, $J^{\mu\nu}(r)=\delta_{\mu\nu}J(r)$
the relevant solution of eq. (\ref{TGG}) has the
form $q_{SG}({\bf S}_1,{\bf S}_2)={\bf S}_1\cdot{\bf S}_2$.
Indeed, the $Q_i$'s may be replaced by unity because of symmetry
and the remaining integrals over ${\bf S}_3$ and ${\bf S}_4$ are
proportional to ${\bf S}_1$ and ${\bf S}_2$, respectively.
This leads to the condition
\begin{equation}\label{Langevin}
c\sum_r \left\{ \frac{\int d{\bf S}\,S^z
\exp(\beta J(r)S^z)}
 {\int d{\bf S}
\exp(\beta J(r)S^z)}\right\}^2=c\sum_r \left\{
\frac{{\rm I}_{\frac{m}{2}}(\beta J(r))}
{{\rm I}_{\frac{m-2}{2}}(\beta J(r))}\right\}^2=1 ,
\end{equation}
where ${\rm I_\nu}$ is the modified Bessel function of index $\nu$.
It is the generalization of the relation
$c\tau_2\equiv c\sum_r \tanh^2\beta J(r)=1$ for Ising spins ($m=1$),
derived in section \ref{ViBr}. For Heisenberg spins $(m=3)$ eq.
(\ref{Langevin})
 becomes
\begin{equation}\label{TGHeis}
c\sum_r \left\{{\rm cotanh}\,\beta J(r)-\frac{1}{\beta J(r)}\right\}^2=1.
\end{equation}
As it involves the classical Langevin function for our
classical spins, one expects it to
contain to the quantum Langevin function for quantum spins.
Eq. (\ref{TGHeis}) is quite natural, as the same mean field form,
but without taking the square,
is known to occur for the ferromagnetic transition.

Eq. (\ref{TGHeis}) again leads to concentration scaling $T_G\sim c$
for RKKY couplings. We have tested it for the situation considered
by Matsubara and Iguchi.\cite{MatIg}
These authors consider a fcc lattice with lattice constant $a_0$ and
with $c=5\%$ spins present. The RKKY coupling
$J(r)=J_0\cos(2k_Fr)(a_0/r)^3$ involves the parameter $k_F=4.91/a_0$;
$J(r)$ is set to zero for $r\ge 3a_0$.
The simulations yield a spin glass transition at $T_g/J_0=0.068\pm0.008$.
We find from the above relation that $T_g/J_0=0.306$.
It shows that $c=5\%$ is not a small concentration, and
that higher order corrections in $c$ are important.


\section{Spin glass behavior at low temperatures}
\label{erg}
In order to see whether spin glass behavior in the form of
breaking of replica symmetry is likely to
occur, we consider the simplest generalization of the SK-model.
It is given by eq. (\ref{Hmf}), (\ref{Phn=}), and \ref{Xl2oc=}).
The model represents
the site-diluted spin system in a Gaussian approximation.

Consider the low temperature behavior of the
replica symmetric solution $M_\alpha=M,\,
q_{\alpha\alpha}=q_d,\,q_{\alpha\neq\beta}=q$.
For small $T$ we observe the ``groundstate dominance'' $s_\alpha^{(j)}
=s_\alpha^{(1)}$ for all $j=1,\cdots,\ell$. This implies
that repeated spin sums reduce to single spin sums. It
leads to a free energy
\BEA \label{bFFFF}
\beta F&=&\frac{T}{2c\hat J(0)}m^2+\frac{1}{2c}\int\dk\left\{
\ln(1-c\beta\hat J(k)(q_d-q))
-\frac{c\beta\hat J(k)q}{1-c\beta\hat J(k)(q_d-q)}
\right\}\nn
&+&\frac{1}{2}(p_dq_d-pq)-\sum_\ell\gamma_\ell \int g(x)\d x
\ln{ 2\cosh {\ell(h+m+x\sqrt{p})}},
\EEA
where $g(x)=\exp(-x^2/2)/\sqrt{2\pi}$ is the Gaussian weight.
It can be checked that $q_d=1-c$.
The other mean field equations are
\begin{eqnarray}
M&\equiv&\frac{Tm}{c\hat J(0)}=\int g(x)dx\sum_\ell \gamma_\ell
\ell \tanh {\ell(h+m+x\sqrt{p})},
\nonumber\\
q&=&\int g(x)dx\sum_\ell \gamma_\ell
\ell^2\tanh^2 {\ell(h+m+x\sqrt{p})},\nn
p&=&\int\dk \frac{c\beta^2 \hat J^2(k)}{(1-c\beta\hat J(k)(q_d-q))^2}, \nn
p_d&=&p+\int\dk \frac{\beta \hat J(k)}{1-c\beta\hat J(k)(q_d-q)} .
\end{eqnarray}
For small $T$ it holds that $p\sim T^{-2}$ is large. One therefore finds
\begin{equation}\label{q-q0}
q=q_d-\sqrt{\frac{2}{\pi}}\sum_\ell\gamma_\ell\frac{\ell^2}
{\ell \sqrt{p}} = q_d-\sqrt{\frac{2}{\pi p}}.
\end{equation}
Note that this result also holds for $c>1/2$, where the $\ell$-sum
is formally divergent.

According to the analysis by de Almeida and Thouless~\cite{AT},
the instability of the replica-symmetric
solution is caused by a negative eigenvalue of the fluctuation matrix.
This mode has been termed ``replicon''.\cite{BM79}
We have proposed the name ``ergodon'' for it, since this name relates to
the underlying physical mechanism of ergodicity breaking,
rather than to one of the mathematical ways to study it.
{}~\cite{Nqsg} The ``mass'' of the
 ``ergodon'' or ``replicon''
 can be derived along the lines of AT. One expands the free energy up to
 second order in $\delta p_\ab$, $\delta q_\ab$, imposing the condition
 $\sum_\alpha \delta p_\ab=\sum_\alpha \delta q_\ab=0$. This finally leads to a
 2 times 2 matrix, which has one large eigenvalue (not of interest to us) and
 an eigenvalue that may become zero and negative. The latter eigenvalue
 has in case of ground state dominance the approximate form
\begin{equation}\label{ergod}
\Lambda(T)\approx 1-
\frac{p}{q}
\sum_\ell\gamma_\ell\int g(x)\d x \frac{\ell^4}{\cosh^4 \ell x\sqrt{p}} .
\end{equation}
Using $q\to q_d=1-c$, this leads to
\begin{equation}
\Lambda(T)\approx 1-(1-2c)\sqrt{\frac{8p}{9\pi}}.
\end{equation}
As $p\sim T^{-2}$, it follows that $\Lambda\sim-(1-2c)/T$,
showing the expected AT instability for $0<c<1/2$. However,
for $c>\frac{1}{2}$
the replica symmetric mode is seen to stabilize at low $T$.

 To see whether the replica symmetric state has correct physics for $c>1/2$,
we consider the zero-point entropy. We take $m=h=0$.
To  order $T^0$ the expression  (\ref{bFFFF}) now becomes
\BEA \label{bFFFF2}
\beta F=\frac{1}{2c}\int\dk\left\{
\ln(1-c\beta\hat J(k)(q_d-q))
-\frac{c\beta\hat J(k)q}{1-c\beta\hat J(k)(q_d-q)}
\right\}-\sqrt{\frac{2p}{\pi}} .
\EEA
It follows that the entropy has the form
\BEQ
S=\frac{1}{2c}\int\dk\left\{
-\ln(1-c\beta\hat J(k)(q_d-q))
-\frac{c\beta\hat J(k)(q_d-q)}{1-c\beta\hat J(k)(q_d-q)}\right\}
< 0 .\EEQ
For $T\to 0$ this goes to a finite but negative value. The same behavior
occurs in the
replica symmetric solution of the SK model, which follows as a limiting
case of the present model.

Thus replica symmetry must be broken for all $c$  at low $T$.
There is a glassy phase, possibly coexisting with (anti-) ferromagnetism.
This is somewhat surprising when the
replica symmetric solution is stable, but similar behavior  was found
in a neural network.\cite{Dots} The present calculation suggests that
infinite order replica symmetry breaking occurs at low $T$. Due
to the ``ground state dominance'', it could be
analyzed along the lines familiar from the SK model.

\section{Ginzburg-Landau expansion near the spinglass transition}
\label{GLstuk}

We use a Ginzburg-Landau expansion of the free energy to investigate which
 kinds of glassy phases can occur in site-disordered systems. The expansion
is derived using results of section~\ref{2.1}. Since it includes
only the lowest order overlap $\langle \sigma_{\alpha}\sigma_{\beta}
\rangle$, it has the same form as the Ginzburg-Landau expansion for the
 SK-model \cite{FischerHertz}.
The difference lies in the prefactors, which now depend on
the concentration of spins and on temperature. We show that these
 prefactors can change sign, giving rise to phase-transitions between
different glassy phase. The phase
diagram of the model is studied as a function of the prefactors.

In order to obtain the Ginzburg-Landau potential, we expand the free
energy eq.~(\ref{Hmf}) in the off-diagonal elements of
$p_{\alpha\beta}$
 and $q_{\alpha\beta}$, which vanish at the transition. In the following,
we redefine $q_{\alpha\alpha}=p_{\alpha\alpha}=0$ and keep track of the
diagonal elements $p_d$ and $q_d$ explicitly. We obtain
\begin{eqnarray}
 \beta F_n & = &
\frac{n}{2c (2\pi)^d}\int {\rm d}^d k \ln(1-c\beta\hat{J}(k)
q_d) -  \frac{1}{2}\sum_{j=2}^{\infty} \frac{1}{j}
c^{j-1}\beta^j J_j \sum_{\alpha}(q^j)_{\alpha\alpha}
 \nonumber \\
 & & +\frac{1}{2}
\sum_{\alpha\beta}q_{\alpha\beta}p_{\alpha\beta}+\frac{n}{2}q_dp_d
\nonumber \\
 & & +n\log 2+\sum_{l=1}^{\infty}\gamma_l\left(1-{\rm tr}_s^{(l)}
\sum_{j=0}^{\infty}
\frac{1}{j!}\left(\frac{1}{2}\sum_{\alpha\beta}p_{\alpha\beta}\sigma_{\alpha}
\sigma_{\beta}\right)^j \exp\left[\frac{1}{2}p_d\sum_{\alpha}
\sigma_{\alpha}^2\right] \right),
\label{Fn pab qab}
\end{eqnarray}
where $J_j$ is the $j$-th moment of the effective coupling (\ref{Jeff})
\BEQ
J_j=\int\frac{{\rm d}^d k}{(2\pi)^d} \hat{J}_{\rm eff}(k)^j.
\EEQ
As is discussed in section~\ref{2.1}, this coupling includes the effects of
clustering of spins which become important at higher spin densities. The
saddle point equations for $p_{\alpha\beta}$ and $q_{\alpha\beta}$
are readily obtained from eq.(\ref{Fn pab qab}).
\BEA
p_{\alpha\beta} & = & \sum_{j=2}^{\infty}\beta^jc^{j-1}J_j
 (q^{j-1})_{\alpha\beta},
\nonumber \\
q_{\alpha\beta} & = & \sum_l\gamma_l{\rm
tr}_s^{(l)}\sigma_{\alpha}\sigma_{\beta}\exp X^{(l)} =
 \langle \sigma_{\alpha}\sigma_{\beta}\rangle .
\label{s.p.eqs} \\
\EEA
Using these equations, the $p_{\alpha\beta}$ are eliminated from the
free energy. We subtract the lowest order contribution, which gives
the paramagnetic background, and rescale
$c\beta^2J_2q_{\alpha\beta}\rightarrow
 q_{\alpha\beta}$. eq.(\ref{Fn pab qab}) now reads
\BEA
\beta F_n & = &
\frac{1}{2}\sum_{j=2}^{\infty}\left(1-\frac{1}{j}\right)
\frac{T^jJ_j}{J_2^j}\sum_{\alpha}(q^j)_{\alpha\alpha} + \tilde{\Phi}_n,
\label{Fn 2}
\EEA
where
\BEQ
\tilde{\Phi}_n=\sum_{l=1}^{\infty}\gamma_l\left(1-{\rm tr}_s^{(l)}
\sum_{j=0}^{\infty}
\frac{1}{j!}\left(\frac{1}{2}\sum_{\alpha\beta}\left(q_{\alpha\beta}
+ \sum_i T^i\frac{J_{2+i}}{J_2^{i+1}} \left(q^{i+1}\right)_{\alpha\beta}
\right)\sigma_{\alpha}
\sigma_{\beta}\right)^j \exp\left[\frac{1}{2}p_d
\sum_{\alpha}\sigma_{\alpha}^2\right] \right)
\label{phi_n expanded}
\EEQ

We expand the free energy to fourth order. $\tilde{\Phi}_n$ gives
the following second-order contribution
\BEQ
\frac{1}{8}\sum_l\gamma_l{\rm tr}_s^{(l)}\sum_{\alpha_1\beta_1\alpha_2\beta_2}
q_{\alpha_1\beta_1}q_{\alpha_2\beta_2}\sigma_{\alpha_1}\sigma_{\beta_1}
\sigma_{\alpha_2}\sigma_{\beta_2}
\exp\left[\frac{1}{2}p_d\sum_{\alpha}
\sigma_{\alpha}^2\right].
\label{phi2}
\EEQ
The trace only has a non-zero value if the replica-indices
are pair-wise equal. Because $q_{\alpha\alpha}=0$,
$\alpha_1=\beta_1$ and $\alpha_2=\beta_2$ give no contribution.
 Eq.(\ref{phi2})
thus gives
\BEQ
\frac{1}{4}\mu_{22}\sum_{\alpha}(q^2)_{\alpha\alpha} ,
\EEQ
where we have introduced the $l$-spin moments $\mu_{k_1\cdots k_m}$
\BEA
\mu_{k_1\cdots k_m} & = & \sum_l\gamma_l\,{\rm tr}_s^{(l)} \sigma_1^{k_1}\cdots
\sigma_m^{k_m} \exp\left[\frac{1}{2}p_d\sum_{\alpha}\sigma_{\alpha}^2\right]
\nonumber \\
 & = & \sum_l\gamma_l\, m_{k_1}^{(l)}\cdots m_{k_m}^{(l)}
\left(m_0^{(l)}\right)^{n-m} \nonumber \\
 & \stackrel{n\rightarrow 0}{=} & \sum_l\gamma_l\,
 \frac{m_{k1}^{(l)}}{m_0^{(l)}}\cdots\frac{m_{km}^{(l)}}{m_0^{(l)}},
\label{mu's}
\EEA
with $m_k^{(l)}={\rm tr}_{\sigma} \sigma^k \exp[p_d \sigma^2/2]$.

The third-order contribution from $j=3$ in eq.(\ref{phi_n expanded}) is
derived in the same way. The replica-indices can be paired up in eight
ways, each yielding a term $\sum_{\alpha}(q^3)_{\alpha\alpha}$.
{}From $j=2$ comes a
term of the same form.

Both $j=2$ and $j=3$ give the fourth order term $\sum_{\alpha\alpha}
(q^4)_{\alpha\alpha}$, $j=4$ gives:
\BEA
 & & \sum_{\alpha\beta}q^4_{\alpha\beta} \; \; \mbox{8 ways} .\nonumber \\
 & & \sum_{\alpha\beta\gamma}\left.q^2_{\alpha\beta}q^2_{\alpha\gamma}
\right|_{\beta\neq\gamma}
\; \; \mbox{48 ways} ,\nonumber \\
 & &
\sum_{\alpha\beta\gamma\delta}\left. q_{\alpha\beta}q_{\beta\gamma}
q_{\gamma\delta}q_{\delta\alpha}\right|_{\alpha\neq\gamma;\beta\neq\delta}
  \; \; \mbox{48 ways} ,\nonumber \\
 & & \sum_{\alpha\beta\gamma\delta}\left.q^2_{\alpha\beta}
q^2_{\gamma\delta}\right|_{\alpha\neq \gamma,\delta ;
\beta\neq \gamma,\delta } \; \; .\mbox{12 ways}
\label{4ths}
\EEA
Since
$(\sum_{\alpha\beta}q^2_{\alpha\beta})^2$ is of order $n^2$,
the last term in eq.~(\ref{4ths}) will not contribute in the limit
$n\rightarrow 0$.
We implement the constraints on the indices by Kronecker delta's and
 then sum them out. For the third term in eq.~(\ref{4ths}),
this for instance gives
\BEA
\sum_{\alpha\beta\gamma\delta}\left. q_{\alpha\beta}q_{\beta\gamma}
q_{\gamma\delta}q_{\delta\alpha}\right|_{\alpha\neq\gamma;\beta\neq\delta}
 & = &
\sum_{\alpha\beta\gamma\delta}
q_{\alpha\beta}q_{\beta\gamma}q_{\gamma\delta}q_{\delta\alpha}
(1-\delta_{\alpha\gamma})(1-\delta_{\beta\delta})
 \nonumber \\
 & = &
\sum_{\alpha\beta\gamma\delta}
q_{\alpha\beta}q_{\beta\gamma}q_{\gamma\delta}q_{\delta\alpha}
-2\sum_{\alpha\beta\gamma}q^2_{\alpha\beta}q^2_{\alpha\gamma}
+\sum_{\alpha\beta}q^4_{\alpha\beta}.
\EEA
Combining these results with the contributions from the first term in
eq.~(\ref{Fn 2}), we finally obtain the following free energy, which
has the same
form as for the SK model.
\begin{eqnarray}
\beta F_n= -\frac{\tau}{2} \sum_{\alpha} (q^2)_{\alpha\alpha}-\frac{w}{6}
\sum_{\alpha}
(q^3)_{\alpha\alpha}-\frac{y_1}{8}\sum_{\alpha\beta}q^4_{\alpha\beta}-
\frac{y_2}{8}\sum_{\alpha\beta\gamma}q^2_{\alpha\beta}q^2_{\alpha\gamma}-
\frac{y_3}{8}\sum_{\alpha}(q^4)_{\alpha\alpha}.
\label{GLfe}
\end{eqnarray}

The prefactors in this expansion are functions of concentration and
temperature. The prefactor of the quadratic term,
$\tau=(\mu_{22}-T^2/cJ_2)/2$, vanishes at the spin glass
temperature $T_g(c)\equiv \sqrt{cJ_2\mu_{22}}$. The other prefactors
in the free energy are given by
\begin{eqnarray}
w & = & \mu_{222}-2\frac{T^3J_3}{cJ_2^3}+3\mu_{22}\frac{TJ_3}{J_2^2} ,\\
y_1 & = & \frac{3}{2}\mu_{2222}+\frac{1}{6}\mu_{44}-\mu_{422} ,\\
y_2 & = & \mu_{422}-3\mu_{2222} ,\\
y_3 & = & \mu_{2222}-3\frac{T^4J_4}{c J_2^4}+
2\mu_{22}\frac{T^2(J_3^2+2J_2J_4)}{J_2^4}+4\mu_{222}\frac{TJ_3}{J_2^2}.
\end{eqnarray}
The paramagnetic
behavior is coded in the parameters $p_d$ and $q_d$, that satisfy the
coupled mean field equations $p_d=\beta J_1$ and $q_d=\mu_2$.
All information on clustering
is contained in $\tau$, $w$, and the $y$'s, so in
the $\mu$'s and the $J_j$.
In the limit $c\to 0$ the $\mu$'s go to unity and for $T\sim \sqrt{c}$
the $J_3$ and $J_4$ terms vanish, so that one recovers
the Ginzburg-Landau free energy of the SK-model.
The important factors then are $w=1$, $y_1=2/3$, while the
values of $y_2$ $(=-2)$ and $y_3$ $(=1)$ are irrelevant.
When following the transition line $T=T_g(c)$ in the $c-T$
phase diagram as function of $c$, it is seen that the higher
$\mu$'s are rapidly oscillating functions.
For instance, if
$J_3$ and $J_4$ are much smaller than $J_2$,
then $y_1$ changes sign
at $c=2.7\%$ and at $c=4.3\%$, while $w$ becomes negative at $6.7\%$.

Based on these observations we are led to assume that the relevant
physics near the phase transition(s) is still contained in the GL
free energy eq.~(\ref{GLfe}). However,
there is no reason to assume that $w$ and $y_1$ will always be positive.
(A sign change of $y_1$ occurs also in a Potts glass.~\cite{GKS})
Given the type of the lattice and the values of the spin-spin couplings, the
$c-T$ phase diagram may exhibit a limited number of
special points ($c_\ast,T_\ast$)
where either $w$ or $y_1$ vanishes, and new phase boundaries originate.
In fig.~\ref{fig1}, a fictitious phase diagram is depicted for a system where
$w$ changes sign at some ($c_\ast,T_\ast$).

In the limit $n\rightarrow 0$, the free energy can be expressed in
terms
 of the Parisi order parameter function $q(x)$ in the usual way
{}~\cite{MPV}.
\begin{eqnarray}
\beta F & = & \int^1_0 {\rm d}x\left\{\frac{\tau}{2}q^2(x)-
\frac{w}{3}q(x)T(x)+\frac{y_1}{8}q^4(x)-\frac{y_2+y_3}{8}q^2(x)\int^1_0{\rm
d}y q^2(y)
+\frac{y_3}{2}T^2(x)\right\},
\label{1}
\end{eqnarray}
where
\begin{eqnarray}
T(x) & = & \frac{1}{2}xq^2(x)+q(x)\int^1_x{\rm d}yq(y)+
\frac{1}{2}\int_0^x{\rm d}yq^2(y) .
\label{T}
\end{eqnarray}

The phasediagram of this model is determined by the parameters $w$ and $y_1$.
The sign of the cubic prefactor determines whether the model exhibits a
second or a first order transition. For positive $y_1$, the low temperature
phase will be replica symmetry broken, for negative $y_1$ replica symmetric.
In the region where $y_1$ is of order $\tau^2$, a sixth order replica
symmetry breaking term becomes relevant. We will discuss this case in
section~\ref{7.3}. The parameters $y_2$ and $y_3$ are only important in the
region where $w\ll 1$, which we will discuss first.

\subsection{Behavior near first order type phase transitions}

Suppose we take $w$ negative and $y_1$ positive. As we lower
temperature,
 the system
undergoes a first order transition to a state which breaks replica symmetry.
At positive $w$, on the other hand, we get a {\em second} order
transition
and the low temperature state will undoubtedly be different from the
one at
 negative $w$ (although it will also break replica symmetry, since we
keep
$y_1$ positive).
We want to study the transition between
these two low-temperature states. We therefore take $y_1>0$ and allow
$w$
to change sign. For $w\sim\sqrt{\tau}\ll 1$, $q$ is also proportional
 to $\sqrt{\tau}$, so
eq.~(\ref{1}) then includes all terms of order $\tau^2$ and should
describe the
behavior near this transition.

At positive $w$, we find the usual transition from the paramagnet
(PM) to a Parisi-type spin glass phase (SG) with infinite replica symmetry
breaking. At $w<0$, however, the order parameter in the
low-temperature phase only has one-step replica symmetry breaking (1RSB).
Solutions of this form have previously been found by Gross,
 Kanter and Sompolinsky \cite{GKS} for the Potts glass and by
Crisanti, Horner and Sommers (CHS) \cite{CHS} for a spherical spin
 glass model with
$p$-spin interaction. We find a continuous transition between the SG and
the 1RSB phase (fig.~\ref{fig3}).

The prefactors $y_2$ and $y_3$, which are only important if $w$ is
small,
 must obey
the following conditions: $y_3<-y_1$ and
$y_2~>~\mbox{max}[0,-2y_1-y_3]$.
 The first
is related to keeping the plateau $q_1$ positive, the second to doing this for
the breakpoint $x_1$ ($q_1$ and $x_1$ are defined below).

\subsubsection{The spin glass (SG I) solution}
First, we need to derive Parisi's infinite-RSB solution in the presence of the
$y_2$ and $y_3$-term. We use the following ansatz for the order
parameter function $q(x)$:
\begin{equation}
q(x)= \left\{ \begin{array}{lcr} q_{nc}(x) & ; & x_0 \leq x \leq x_1 \\
                                 q_1\equiv q_{nc}(x_1) & ; & x_1\leq x \leq 1
\end{array} \right.
\label{2}
\EEQ
Because $q(x)$ is continuous, it is completely determined by the saddle-point
equation $\delta F/\delta q(x) =0$,
\begin{eqnarray}
 &  & \tau q(x) -wT(x)
+\frac{y_1}{2}q^3(x)-\frac{y_2+y_3}{2}q(x)\int^1_0{\rm d}yq^2(y)
\nonumber \\
 &  & +y_3\left[T(x)R(x) +
 \int_0^x{\rm d}yT(y)q(y)+q(x)\int^1_x{\rm d}yT(y)\right] = 0,
\label{3}
\end{eqnarray}
where
\BEQ
R(x)  =   xq(x)+\int_x^1{\rm d}yq(y).
\label{R}
\EEQ
Both $T(x)$ and $R(x)$ are constant on the plateau at $x\geq x_1$.

The stability of the SG solution is determined by the sign of the
``ergodon''-function $\Lambda(x)$, which is an eigenvalue of the
fluctuation-matrix
$\delta^2 F[q]/\delta q(x)\delta q(y)$.
\BEA
\Lambda(x)= \tau - wR(x)+\frac{3y_1}{2}q^2(x)
-\frac{y_2+y_3}{2}\int^1_0{\rm d}yq^2(y)+y_3\left[R^2(x)+xT(x)
+\int_x^1{\rm d}yT(y)\right] .
\EEA
Differentiating the saddle-point equation eq.~(\ref{3}) with respect
to $x$, we obtain $q'(x)
\Lambda(x)=0$.
This implies that the SG solution is marginally stable on the interval
$0\leq x\leq x_1$. Moreover, it is also marginally stable for $x>x_1$, because
$q(x)$ is continuous at the breakpoint. This is of course a well-known feature
of this state.

The fourth derivative of the saddle-point equation with respect to $x$ yields
a second order differential equation for $q_{nc}(x)$. Its second
derivative can be used to fix the height of the plateau $q_1$. We obtain,
\begin{equation}
q_{nc}(x)=\frac{w\sqrt{y_1+y_3 x_1^2}x}{3(y_1+y_3 x_1)\sqrt{y_1+y_3x^2}}.
\label{qnc(x)}
\end{equation}

Finally, the breakpoint $x_1$ is given by the implicit equation
$\Lambda(x_1)=0$.
\begin{equation}
\sigma=\frac{x_1}{3(y_1+y_3 x_1)}-\frac{(3y_1+3y_3-2y_2)x_1^2+\frac{y_2}{y_3}(y_1+y_3 x_1^2)
\left(x_1-\alpha\mbox{atanh}\left(\frac{x_1}{\alpha}\right)\right)}{18(y_1+y_3
x_1)^2},
\end{equation}
where $\sigma = \tau/w^2$ and $\alpha=\sqrt{-y_1/y_3}$. Near the
 transition, this
yields $x_1\simeq 3 y_1 \sigma$.

For $\sigma \ll 1$, $w>0$, eq.(\ref{qnc(x)}) reproduces
Parisi's solution of the SK model with $x_1\ll 1$ (fig.~\ref{fig2}a). When
$x_1$ approaches $\alpha$, $q(x)$ squeezes into a stepfunction,
with the discontinuity located at $x=x_1$ (fig.\ref{fig2}f). This is the
transition from infinite to one-step replica symmetry breaking. It
occurs in the region $w<0$, at $\sigma=\sigma_c=1/6y_3+y_2/18y_3^2(1-\alpha)$
(fig.~\ref{fig3}).

\subsubsection{One-step replica symmetry breaking solutions}
The order parameter in the 1RSB state has the form of a step function
\begin{equation}
q(x)= \left\{ \begin{array}{lcr} 0 & ; & 0 \leq x \leq x_1 \\
                                 q_1  & ; & x_1\leq x \leq 1
\end{array} \right.
\EEQ
Two conditions are needed in order to fix both the plateau $q_1$ and
 the breakpoint
$x_1$. Next to the saddle-point equation $\partial F/\partial
q_1=0$, we may impose either stationarity with respect to the breakpoint,
$\partial F/\partial x_1=0$, or marginal stability of the
ergodon on
one of the plateaus, $\Lambda_{0/1}=0$. We will consider both procedures.
The first gives
the proper static solution (or the solution for an infinitely slow
cooling rate), while the second solution
ought to occur in dynamics.

The saddle-point equation  $\partial F/\partial q_1=0$ reads
\begin{equation}
\tau-w\left(1-\frac{1}{2}x_1\right)q_1+\frac{1}{2}(y_1-y_2(1-x_1)
+3y_3(1-x_1)+y_3x_1^2)q_1^2=0.
\end{equation}
The static solution must satisfy the additional equation
\begin{eqnarray}
-\tau+w\left(1-\frac{2}{3}x_1\right)q_1-\frac{1}{4}y_1q_1^2+\frac{1}{2}y_2
(1-x_1)q_1^2-y_3\left(\frac{3}{2}-x_1+\frac{3}{4}x_1^2\right)q_1^2=0.
\label{dFdx1=0}
\end{eqnarray}
The breakpoint of the marginally stable solution is fixed by
\begin{equation}
\Lambda_1=0 \Leftrightarrow
-\tau+wq_1-\frac{3}{2}(y_1+y_3)q_1^2+\frac{1}{2}y_2(1-x_1)q_1^2=0.
\label{Lambda1=0}
\end{equation}
In principle, the condition $\Lambda_0=0$ could also yield a dynamical
solution, but this 1RSB state turns out to be unstable.

We find the following static solution
\BEQ
q_1^{\rm stat}  =
\frac{wx_1}{\frac{3}{2}y_1+3y_3x_1\left(1-\frac{1}{2}x_1\right)}.
\label{q1 1RSB stat }
\EEQ
This solution sets in from $x_1=1$, at a negative value of $\sigma$,
$\sigma_1^{\rm stat} = 1/9(y_1+y_3)$. The transition is first order
in the sense
that the order-parameter changes discontinuously and that the
transition occurs before the AT instability of the PM phase. There is
no latent heat, however. In fact, there is a problem, because the
continuation of the PM phase has a lower free energy than the
new 1RSB solution. The system nonetheless needs to go to the 1RSB state,
because the PM becomes unstable at $\sigma=0$. This problem was also
encountered by CHS.

The ergodon mass of the static 1RSB is the same on both plateaus
\BEQ
\Lambda_0^{\rm stat}=\Lambda_1^{\rm stat}=
-\frac{1}{4}(y_1+y_3 x_1^2)(q_1^{ \rm stat})^2.
\EEQ
It goes unstable at the transition to the SG phase,
$\sigma=\sigma_c$, where $x_1$ becomes equal to $\alpha$.

The dynamic solution is given by
\BEQ
q_1^{\rm marg}  =  \frac{wx_1}{2y_1+y_3x_1(3-x_1)}.
\label{q1 1RSB marg}
\EEQ
It also sets in a $x_1=1$, but at a higher temperature than the static
1RSB: $\sigma_1^{\rm marg}=1/8(y_1+y_3)$. Close to $\sigma_1^{\rm marg}$, its
free energy is lower than that of the PM ($\Delta F$ even has a finite
slope, but there is no latent heat~\cite{Nmaxmin},\cite{NEhren}).
However, it becomes higher than $F_{\rm PM}$
before the PM goes unstable at $\sigma=0$, so we have the same problem
as with the static solution.

The dynamic solution has $\Lambda_1=0$, while
the other eigenvalue is given by
\BEQ
\Lambda_0^{\rm marg}=-\frac{1}{2}(y_1+y_3 x_1^2)(q_1^{\rm marg})^2.
\EEQ
It vanishes at the transition to the SG phase.

The subject of dynamical transitions has made large progress since the
start of this work, for a review see ~\cite{BCKM}.
It is now understood that the marginality
criterion signals dynamics in the highest, marginal  states.
This is the regime reached when first taking the thermodynamic limit
and then letting time become large.
In realistic systems  one expects to go to lower states in the course
of time, in a way that can be fixed, to some extent,
 by the cooling procedure. That
happens in mean field spin glasses provided one considers dynamics at
time scales exponential in the system size~\cite{Nthermo}.
In other words, the marginality only points at the onset of a more
interesting behavior. Dynamics in marginal states itself
has a very limited meaning; it is a mean field artefact.
The final result of these investigations has been the development of a
picture for the thermodynamics of the non-equilibrium glassy state.
Apart from the real temperature there occurs an effective temperature
$T_e=T/x_1$, at which the system's slow modes are at
quasi-equilibrium~\cite{Nthermo}\cite{Nhammer}.
This approach explains the old paradoxes related to the Ehrenfest relations
and the Prigogine-Defay ratio~\cite{NEhren}, that were the basis for
the general belief that thermodynamics does not work for the glassy state.

\subsection{The transition from the Edwards-Anderson to the spin glass phase}
\label{7.3}
Another region of interest is where the prefactor of the replica symmetry
breaking term, $y_1$, changes sign. For $y_1 \sim -1$, $w\sim 1$, the model is
in the Edwards Anderson (EA) phase, which has a replica symmetric
order parameter (fig.\ref{fig2}e). At positive $y_1$, it is in the SG
phase.  At the transition between these two phases, $y_1\ll 1$
and higher order replica symmetry breaking
terms become relevant. We therefore consider the following free energy
\BEA
\beta F_n= -\frac{\tau}{2} \sum_{\alpha} (q^2)_{\alpha\alpha}-\frac{w}{6}
\sum_{\alpha}
(q^3)_{\alpha\alpha}-\frac{y_1}{8}\sum_{\alpha\beta}q^4_{\alpha\beta}-
\frac{y_5}{8}\sum_{\alpha\beta}q^3_{\alpha\beta}(q^2)_{\alpha\beta}-
\frac{y_6}{6}\sum_{\alpha\beta}q^6_{\alpha\beta}.
\label{Fn 5en6}
\EEA
The $y_2$ and $y_3$ terms are omitted, since they
are not important for $w\sim 1$. We have included the most dangerous
fifth and sixth order term. The full list of fifth order terms is
\BEA & &
-\sum_{\alpha}(q^5)_{\alpha\alpha}\left(\frac{1}{10}\mu_{22222}
+\frac{1}{2}\mu_{22}T^3\frac{J_5J_2+J_3J_4}{J_2^5}
+\frac{1}{2}\mu_{222}T^2\frac{J_4J_2+J_3^2}{J_2^4}
+\frac{1}{2}\mu_{2222}T\frac{J_3}{J_2^2}-\frac{2}{5}\frac{T^5J_5}{c
J_2^5}\right) \nonumber \\ & &
-\sum_{\alpha\beta}q^3_{\alpha\beta}(q^2)_{\alpha\beta}
\left(\frac{3}{4}\mu_{22222}-\frac{1}{2}\mu_{4222}
+\frac{1}{12}\mu_{442}+\frac{1}{2}T
\frac{J_3}{J_2^2}y_1\right)\nonumber \\ & &
-\sum_{\alpha\beta}q^2_{\alpha\beta}(q^3)_{\alpha\alpha}
\left(-\frac{3}{4}\mu_{22222}+\frac{1}{4}\mu_{4222}\right)
-\sum_{\alpha\beta\gamma}q^2_{\alpha\beta}(q^3)_{\gamma\gamma}
\frac{1}{24}\mu_{22222}
\label{fifthorder}\EEA
The last of those is of order $n^2$ and does not contribute for
$n\rightarrow 0$. The first and the third term are also neglected, since they
have a structure similar to the cubic term, but are of higher order in $\tau$.
It turns out that the $y_5$-term (the second of eq.~(\ref{fifthorder}))
can be absorbed into a redefinition of $y_1$
by using the saddle point equation $(q^2)_{\alpha\beta}\simeq -2\tau
q_{\alpha\beta}/w$. We therefore replace $y_1$ by $\tilde{y}_1=y_1-
2\tau y_5/w$ and omit the fifth order term.
The prefactor of the most dangerous sixth order term is given by
\BEA
y_6=-\frac{15}{4}\mu_{222222}+\frac{15}{4}\mu_{42222}-\frac{1}{4}\mu_{6222}
-\frac{15}{16}\mu_{4422}+\frac{1}{8}\mu_{642}-\frac{1}{240}\mu_{66}.
\EEA

We may expect new behavior when
$\tilde{y}_1$ and $y_6$ become of the same order of magnitude.
Writing $\tilde{y}_1=2 y_4 \tau^2/w^2$, we study the phasediagram
as a function
of $y_4$ for $\tau \ll 1$ and fixed. We find the following results:

For $y_6>0$, the model has a SG phase with nonzero lower plateau $q_0$. It
interpolates between the EA phase, where $q_0=q_1$,
and the SG phase, where $q_0=0$ (fig.\ref{fig4}).

For negative $y_6$, there are two
saddle points at the transition between the EA and the SG phase.
The metastable
saddle point consists of a 1RSB solution which sets in from the EA phase,
followed by a transition to a new phase which we call SG{\em IV}
 (fig.\ref{fig2}d).
At larger $y_4$ there is a second transition from SG{\em IV} to SG.

The order parameter of the stable saddle point is shown in
fig.\ref{fig2}c. This
is also a new solution, which we call SG{\em III}. It has a lower free energy
than the 1RSB and the SG{\em IV} solution. The SG{\em III}
sets in from the EA phase with $x_1=q_0=0$,
and evolves into the SG at higher $y_4$, where $q_0$ becomes equal to $q_1$.

The phasediagram for this region of parameter space is shown in
fig.\ref{fig5}.

The solutions in this region are obtained in much the same way as the
SG and the
1RSB. We therefore only show that there is no direct transition from
 the EA to the SG
phase and then simply state the results for the phases which occur
in between.

As $n\rightarrow 0$, the free energy eq.~(\ref{Fn 5en6}) takes the
 following form in
terms of the order parameter function $q(x)$
\begin{eqnarray}
\beta F & = & \int^1_0 {\rm d}x\left\{\frac{\tau}{2}q^2(x)-
\frac{w}{3}q(x)T(x)+\frac{y_4\tau^2}{4 w^2}q^4(x)
-\frac{y_6}{6}q^6(x)\right\}.
\label{F y6}
\end{eqnarray}

Taking a replica symmetric ansatz, we now find for the order parameter
in the EA phase
\BEA
q_{\rm EA} = \frac{\tau}{w}+\frac{y_4-y_6}{w^5}\tau^4+{\cal O}(\tau^5).
\EEA

The SG solution is again determined by the saddle point equation
 $\delta F/\delta q(x)=0$.
{}From the second derivative of this equation, the non-constant part of $q(x)$
is obtained
\BEQ
x(q)=\frac{1}{w}\left(6y_4q\frac{\tau^2}{w^2}-20y_6q^3\right).
\label{x(q)}
\EEQ
The upper plateau and the breakpoint of the SG order parameter are given by
\BEA
q_1 & = & \frac{\tau}{w}+\frac{3y_4-5y_6}{w^5}\tau^4 +{\cal O}(\tau^5),
\nonumber  \\
x_1 & = &  x(q_1)=\frac{6y_4-20y_6}{w^4}\tau^3 +{\cal O}(\tau^4).
\EEA

{}From these equations, it is seen that there is no smooth transition
 from the EA
to the SG phase for nonzero $y_6$. The details of the various solutions which
occur in the region $|y_4|\sim |y_6|$
are listed below. For all solution of the SG-type, the non-constant part of
the order-parameter is given by eq.(\ref{x(q)}).

\underline{SG with $q_0\neq 0$}
\BEA q_0 & = &  \frac{\tau}{w}\sqrt{\frac{y_4}{2 y_6}} +{\cal
O}(\tau^2),
 \nonumber \\
 q_1 & = &  q_{1 SG} \; \; ; \; \; x_1=x_{1 SG}.
\EEA

\underline{SG{\em III}}
\begin{eqnarray}
q_0 & = & -\frac{\tau}{4w}+\frac{\tau}{4w}\sqrt{1+\frac{8(y_4-y_6)}{3y_6}}
+{\cal O}(\tau^2)  \\
q_1 & = & \frac{1}{w}\tau+\frac{y_4}{w^4}(6q_0\tau^3-3wq_0^2\tau^2)
-\frac{y_6}{w^2}(20q_0^3\tau-15wq_0^4) +{\cal O}(\tau^5)
\end{eqnarray}

\underline{1RSB}
\BEA
q_1 & = & \frac{\tau}{w}+\frac{5 y_4-6 y_6}{2w^5}\tau^4 +{\cal O}(\tau^5),
\nonumber \\
x_1 & = & \frac{3 y_4- 4 y_6}{w^4}\tau^3.
\EEA

\underline{SG{\em IV}}
\begin{eqnarray}
q_0 & = & -\frac{\tau}{2w}+\frac{\tau}{2w}\sqrt{1+\frac{y_4-2y_6}{y_6}}
+{\cal O}(\tau^2) \nonumber \\
q_1 & = &
\frac{\tau}{w}+2\frac{y_4-y_6}{w^5}(\tau^4+wq_0\tau^3)
-\frac{y_4+2y_6}{w^3}q_0^2\tau^2-\frac{y_6}{w^2}(2q_0^3\tau-3wq_0^4)
+{\cal O}(\tau^5) \\
x_1 & = &
2\frac{y_4-y_6}{w^4}(\tau^3+2wq_0\tau^2)-\frac{y_6}{w^2}(6q_0^2\tau+8wq_0^3)
+{\cal O}(\tau^4) \nonumber
\end{eqnarray}

\subsection{Unsuccessful attempts}
Since the procedure for finding saddle points of the free energy is to
guess an Ansatz
and then to check if it works, we cannot guarantee that we have found
 everything
there is. In particular, we have not found satisfactory solutions for
 the region
$w<0$ and $y_1<0$. Also, there may be extra solutions in the regions where we
have already found stable saddle points, as the results for $y_1\ll 1$ show.
There are however some states which we definitely know {\em not} to be there,
since we have unsuccessfully tried them. They are shown in
fig.~\ref{fig6}. Most
notably, we have not found a stable 2RSB state (fig.~\ref{fig6}a)
for any values of the parameters.
One-step and infinite replica symmetry breaking seem to be the only types of
RSB that occur. Also 1RSB states with a non-zero lower plateau (in zero
magnetic field) have been tried without success. States $b$,  $c$ and $d$
 were thought to evolve from the 1RSB at small $w$, but none of them
do. State $b$ (SG{\em II}) does occur in the Potts glass \cite{GKS}.

\vspace{1cm}
\begin{figure}
\centering
\begin{picture}(0,0)%
\includegraphics{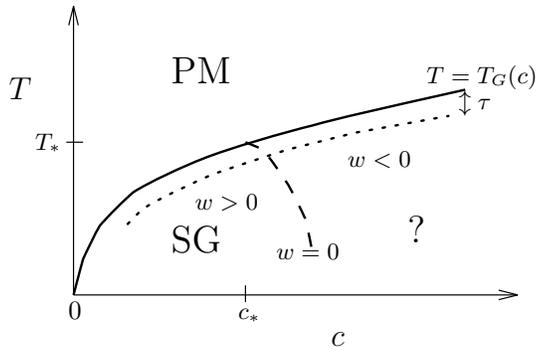}%
\end{picture}%
\setlength{\unitlength}{0.012500in}
\begingroup\makeatletter\ifx\SetFigFont\undefined
\def\x#1#2#3#4#5#6#7\relax{\def\x{#1#2#3#4#5#6}}%
\expandafter\x\fmtname xxxxxx\relax \def\y{splain}%
\ifx\x\y   
\gdef\SetFigFont#1#2#3{%
  \ifnum #1<17\tiny\else \ifnum #1<20\small\else
  \ifnum #1<24\normalsize\else \ifnum #1<29\large\else
  \ifnum #1<34\Large\else \ifnum #1<41\LARGE\else
     \huge\fi\fi\fi\fi\fi\fi
  \csname #3\endcsname}%
\else
\gdef\SetFigFont#1#2#3{\begingroup
  \count@#1\relax \ifnum 25<\count@\count@25\fi
  \def\x{\endgroup\@setsize\SetFigFont{#2pt}}%
  \expandafter\x
    \csname \romannumeral\the\count@ pt\expandafter\endcsname
    \csname @\romannumeral\the\count@ pt\endcsname
  \csname #3\endcsname}%
\fi
\fi\endgroup
\begin{picture}(214,142)(40,638)
\put( 40,741){\makebox(0,0)[lb]{\smash{\SetFigFont{12}{14.4}{rm}$T$}}}
\put(225,737){\makebox(0,0)[lb]{\smash{\SetFigFont{9}{10.8}{rm}
$\updownarrow \tau$}}}
\put(215,748){\makebox(0,0)[lb]{\smash{\SetFigFont{9}{10.8}{rm}$T=T_G(c)$}}}
\put(118,695){\makebox(0,0)[lb]{\smash{\SetFigFont{9}{10.8}{rm}$w>0$}}}
\put( 51,720){\makebox(0,0)[lb]{\smash{\SetFigFont{9}{10.8}{rm}$T_{\ast}$}}}
\put(182,713){\makebox(0,0)[lb]{\smash{\SetFigFont{9}{10.8}{rm}$w<0$}}}
\put(207,681){\makebox(0,0)[lb]{\smash{\SetFigFont{14}{16.8}{rm}?}}}
\put( 65,649){\makebox(0,0)[lb]{\smash{\SetFigFont{10}{12.0}{rm}0}}}
\put(108,677){\makebox(0,0)[lb]{\smash{\SetFigFont{14}{16.8}{rm}SG}}}
\put(108,748){\makebox(0,0)[lb]{\smash{\SetFigFont{14}{16.8}{rm}PM}}}
\put(175,638){\makebox(0,0)[lb]{\smash{\SetFigFont{12}{14.4}{rm}$c$}}}
\put(152,674){\makebox(0,0)[lb]{\smash{\SetFigFont{9}{10.8}{rm}$w=0$}}}
\put(136,649){\makebox(0,0)[lb]{\smash{\SetFigFont{9}{10.8}{rm}$c_{\ast}$}}}
\end{picture}
\caption{ $c-T$ phase diagram for a fictitious system with
a line $w(c,T)=0$. PM=paramagnet; SG=spin glass.}
\label{fig1}
\end{figure}

\begin{figure}
\centering
\begin{picture}(0,0)%
\includegraphics{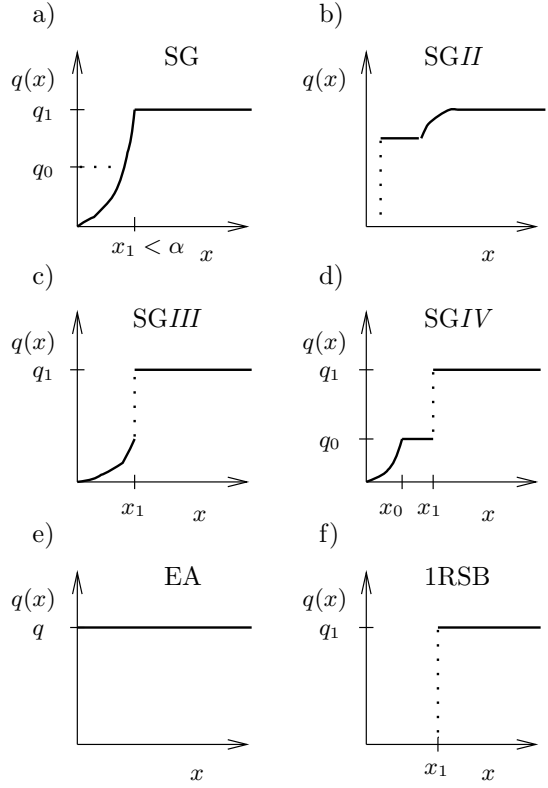}%
\end{picture}%
\setlength{\unitlength}{0.01250000in}%
\begingroup\makeatletter\ifx\SetFigFont\undefined
\def\x#1#2#3#4#5#6#7\relax{\def\x{#1#2#3#4#5#6}}%
\expandafter\x\fmtname xxxxxx\relax \def\y{splain}%
\ifx\x\y   
\gdef\SetFigFont#1#2#3{%
  \ifnum #1<17\tiny\else \ifnum #1<20\small\else
  \ifnum #1<24\normalsize\else \ifnum #1<29\large\else
  \ifnum #1<34\Large\else \ifnum #1<41\LARGE\else
     \huge\fi\fi\fi\fi\fi\fi
  \csname #3\endcsname}%
\else
\gdef\SetFigFont#1#2#3{\begingroup
  \count@#1\relax \ifnum 25<\count@\count@25\fi
  \def\x{\endgroup\@setsize\SetFigFont{#2pt}}%
  \expandafter\x
    \csname \romannumeral\the\count@ pt\expandafter\endcsname
    \csname @\romannumeral\the\count@ pt\endcsname
  \csname #3\endcsname}%
\fi
\fi\endgroup
\begin{picture}(226,330)(35,428)
\put( 35,503){\makebox(0,0)[lb]{\smash{\SetFigFont{9}{10.8}{rm}$q(x)$}}}
\put( 44,491){\makebox(0,0)[lb]{\smash{\SetFigFont{9}{10.8}{rm}$q$}}}
\put(157,611){\makebox(0,0)[lb]{\smash{\SetFigFont{9}{10.8}{rm}$q(x)$}}}
\put(232,539){\makebox(0,0)[lb]{\smash{\SetFigFont{9}{10.8}{rm}$x$}}}
\put(164,599){\makebox(0,0)[lb]{\smash{\SetFigFont{9}{10.8}{rm}$q_1$}}}
\put(205,542){\makebox(0,0)[lb]{\smash{\SetFigFont{9}{10.8}{rm}$x_1$}}}
\put(189,542){\makebox(0,0)[lb]{\smash{\SetFigFont{9}{10.8}{rm}$x_0$}}}
\put(164,570){\makebox(0,0)[lb]{\smash{\SetFigFont{9}{10.8}{rm}$q_0$}}}
\put(164,638){\makebox(0,0)[lb]{\smash{\SetFigFont{10}{12.0}{rm}d)}}}
\put(157,721){\makebox(0,0)[lb]{\smash{\SetFigFont{9}{10.8}{rm}$q(x)$}}}
\put(164,748){\makebox(0,0)[lb]{\smash{\SetFigFont{10}{12.0}{rm}b)}}}
\put(157,503){\makebox(0,0)[lb]{\smash{\SetFigFont{9}{10.8}{rm}$q(x)$}}}
\put(232,430){\makebox(0,0)[lb]{\smash{\SetFigFont{9}{10.8}{rm}$x$}}}
\put(208,433){\makebox(0,0)[lb]{\smash{\SetFigFont{9}{10.8}{rm}$x_1$}}}
\put(164,491){\makebox(0,0)[lb]{\smash{\SetFigFont{9}{10.8}{rm}$q_1$}}}
\put(232,648){\makebox(0,0)[lb]{\smash{\SetFigFont{9}{10.8}{rm}$x$}}}
\put( 44,530){\makebox(0,0)[lb]{\smash{\SetFigFont{10}{12.0}{rm}e)}}}
\put(164,530){\makebox(0,0)[lb]{\smash{\SetFigFont{10}{12.0}{rm}f)}}}
\put( 86,620){\makebox(0,0)[lb]{\smash{\SetFigFont{10}{12.0}{rm}SG{\em III}}}}
\put(208,730){\makebox(0,0)[lb]{\smash{\SetFigFont{10}{12.0}{rm}SG{\em II}}}}
\put(208,620){\makebox(0,0)[lb]{\smash{\SetFigFont{10}{12.0}{rm}SG{\em IV}}}}
\put( 99,512){\makebox(0,0)[lb]{\smash{\SetFigFont{10}{12.0}{rm}EA}}}
\put( 99,730){\makebox(0,0)[lb]{\smash{\SetFigFont{10}{12.0}{rm}SG}}}
\put(208,512){\makebox(0,0)[lb]{\smash{\SetFigFont{10}{12.0}{rm}1RSB}}}
\put( 35,611){\makebox(0,0)[lb]{\smash{\SetFigFont{9}{10.8}{rm}$q(x)$}}}
\put(111,539){\makebox(0,0)[lb]{\smash{\SetFigFont{9}{10.8}{rm}$x$}}}
\put( 81,542){\makebox(0,0)[lb]{\smash{\SetFigFont{9}{10.8}{rm}$x_1$}}}
\put( 44,599){\makebox(0,0)[lb]{\smash{\SetFigFont{9}{10.8}{rm}$q_1$}}}
\put( 44,638){\makebox(0,0)[lb]{\smash{\SetFigFont{10}{12.0}{rm}c)}}}
\put( 35,721){\makebox(0,0)[lb]{\smash{\SetFigFont{9}{10.8}{rm}$q(x)$}}}
\put( 77,651){\makebox(0,0)[lb]{\smash{\SetFigFont{9}{10.8}{rm}$x_1<\alpha$}}}
\put(114,648){\makebox(0,0)[lb]{\smash{\SetFigFont{9}{10.8}{rm}$x$}}}
\put( 44,708){\makebox(0,0)[lb]{\smash{\SetFigFont{9}{10.8}{rm}$q_1$}}}
\put( 44,748){\makebox(0,0)[lb]{\smash{\SetFigFont{10}{12.0}{rm}a)}}}
\put( 44,684){\makebox(0,0)[lb]{\smash{\SetFigFont{9}{10.8}{rm}$q_0$}}}
\put(110,430){\makebox(0,0)[lb]{\smash{\SetFigFont{9}{10.8}{rm}$x$}}}
\end{picture}
\vspace{.2 cm}
\caption{The order-parameter function $q(x)$ for various phases. The SG{\em II}
solution has been found for the Potts glass by Gross {em et. al.},
 but does not occur here.}
\label{fig2}
\end{figure}

\vspace{1cm}
\begin{figure}
\centering
\begin{picture}(0,0)%
\includegraphics{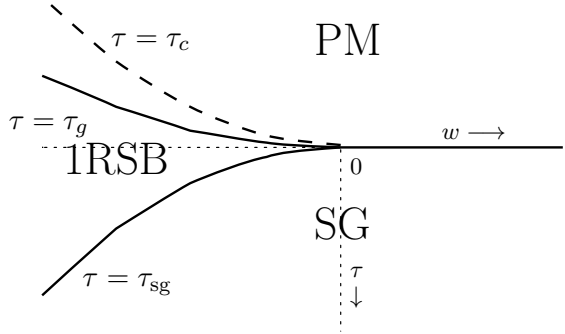}%
\end{picture}%
\setlength{\unitlength}{0.012500in}
\begingroup\makeatletter\ifx\SetFigFont\undefined
\def\x#1#2#3#4#5#6#7\relax{\def\x{#1#2#3#4#5#6}}%
\expandafter\x\fmtname xxxxxx\relax \def\y{splain}%
\ifx\x\y   
\gdef\SetFigFont#1#2#3{%
  \ifnum #1<17\tiny\else \ifnum #1<20\small\else
  \ifnum #1<24\normalsize\else \ifnum #1<29\large\else
  \ifnum #1<34\Large\else \ifnum #1<41\LARGE\else
     \huge\fi\fi\fi\fi\fi\fi
  \csname #3\endcsname}%
\else
\gdef\SetFigFont#1#2#3{\begingroup
  \count@#1\relax \ifnum 25<\count@\count@25\fi
  \def\x{\endgroup\@setsize\SetFigFont{#2pt}}%
  \expandafter\x
    \csname \romannumeral\the\count@ pt\expandafter\endcsname
    \csname @\romannumeral\the\count@ pt\endcsname
  \csname #3\endcsname}%
\fi
\fi\endgroup
\begin{picture}(237,140)(40,637)
\put(187,649){\makebox(0,0)[lb]{\smash{\SetFigFont{9}{10.8}{rm}$\downarrow$}}}
\put(187,660){\makebox(0,0)[lb]{\smash{\SetFigFont{9}{10.8}{rm}$\tau$}}}
\put(187,703){\makebox(0,0)[lb]{\smash{\SetFigFont{9}{10.8}{rm}0}}}
\put( 83,757){\makebox(0,0)[lb]{\smash{\SetFigFont{11}{13.2}{rm}
$\tau=\tau_c$}}}
\put(172,753){\makebox(0,0)[lb]{\smash{\SetFigFont{17}{20.4}{rm}PM}}}
\put(222,718){\makebox(0,0)[lb]{\smash{\SetFigFont{9}{10.8}{rm}
$w\longrightarrow $}}}
\put(172,676){\makebox(0,0)[lb]{\smash{\SetFigFont{17}{20.4}{rm}SG}}}
\put( 71,656){\makebox(0,0)[lb]{\smash{\SetFigFont{11}{13.2}{rm}
$\tau=\tau_{\rm sg}$}}}
\put( 67,703){\makebox(0,0)[lb]{\smash{\SetFigFont{17}{20.4}{rm}1RSB}}}
\put( 40,722){\makebox(0,0)[lb]{\smash{\SetFigFont{11}{13.2}{rm}
$\tau=\tau_g$}}}
\end{picture}
\vspace{.5cm}
\caption{ $\tau-w$ phase diagram for a system with
$y_1>0$, $y_3<-y_1$ and $y_2~>~\mbox{max}[0,-2y_1-y_3]$ ; with $w$ increasing
from right to left
 it may appear in Fig. 1 around the point ($c_\ast$, $T_\ast$).}
\label{fig3}
\end{figure}

\begin{figure}
\centering
\begin{picture}(0,0)%
\includegraphics{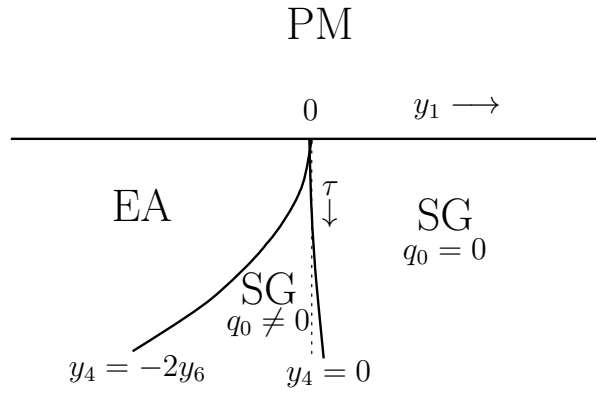}%
\end{picture}%
\setlength{\unitlength}{0.012500in}%
\begingroup\makeatletter\ifx\SetFigFont\undefined
\def\x#1#2#3#4#5#6#7\relax{\def\x{#1#2#3#4#5#6}}%
\expandafter\x\fmtname xxxxxx\relax \def\y{splain}%
\ifx\x\y   
}\gdef\SetFigFont#1#2#3{%
  \ifnum #1<17\tiny\else \ifnum #1<20\small\else
  \ifnum #1<24\normalsize\else \ifnum #1<29\large\else
  \ifnum #1<34\Large\else \ifnum #1<41\LARGE\else
     \huge\fi\fi\fi\fi\fi\fi
  \csname #3\endcsname}%
\else
\gdef\SetFigFont#1#2#3{\begingroup
  \count@#1\relax \ifnum 25<\count@\count@25\fi
  \def\x{\endgroup\@setsize\SetFigFont{#2pt}}%
  \expandafter\x
    \csname \romannumeral\the\count@ pt\expandafter\endcsname
    \csname @\romannumeral\the\count@ pt\endcsname
  \csname #3\endcsname}%
\fi
\fi\endgroup
\begin{picture}(245,164)(44,616)
\put(208,669){\makebox(0,0)[lb]{\smash{\SetFigFont{12}{14.4}{rm}$q_0=0$}}}
\put(140,650){\makebox(0,0)[lb]{\smash{\SetFigFont{17}{20.4}{rm}SG}}}
\put( 86,685){\makebox(0,0)[lb]{\smash{\SetFigFont{17}{20.4}{rm}EA}}}
\put(159,761){\makebox(0,0)[lb]{\smash{\SetFigFont{17}{20.4}{rm}PM}}}
\put(208,730){\makebox(0,0)[lb]{\smash{\SetFigFont{12}{14.4}{rm}
$y_1\longrightarrow$}}}
\put(134,639){\makebox(0,0)[lb]{\smash{\SetFigFont{12}{14.4}{rm}$q_0\neq 0$}}}
\put(166,727){\makebox(0,0)[lb]{\smash{\SetFigFont{12}{14.4}{rm}0}}}
\put( 68,620){\makebox(0,0)[lb]{\smash{\SetFigFont{12}{14.4}{rm}$y_4=-2y_6$}}}
\put(214,681){\makebox(0,0)[lb]{\smash{\SetFigFont{17}{20.4}{rm}SG}}}
\put(174,685){\makebox(0,0)[lb]{\smash{\SetFigFont{12}{14.4}{rm}$\downarrow$}}}
\put(174,696){\makebox(0,0)[lb]{\smash{\SetFigFont{12}{14.4}{rm}$\tau$}}}
\put(159,618){\makebox(0,0)[lb]{\smash{\SetFigFont{12}{14.4}{rm}$y_4=0$}}}
\end{picture}
\vspace{.5cm}
\caption{$y_1-\tau$ phase diagram for $w>0$, $ y_6>0$.
The 
function $q(x)$ in the SG phase
is drawn in figure 2.a for the case $q_0=0$.
In the EA-phase $q(x)$ is constant (no RSB).}
\label{fig4}
\vspace{.5cm}
\end{figure}

\begin{figure}
\centering
\begin{picture}(0,0)%
\includegraphics{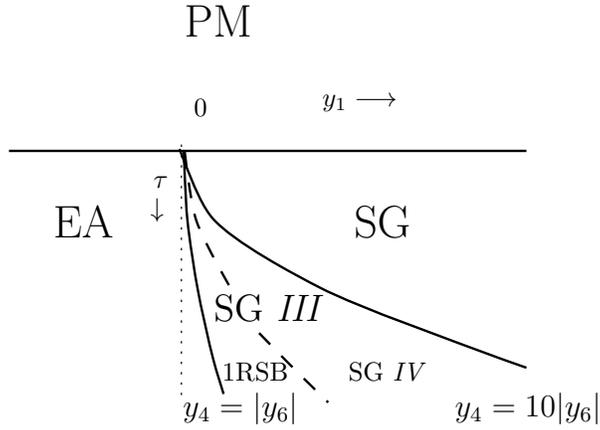}%
\end{picture}%
\setlength{\unitlength}{0.012500in}%
\begingroup\makeatletter\ifx\SetFigFont\undefined
\def\x#1#2#3#4#5#6#7\relax{\def\x{#1#2#3#4#5#6}}%
\expandafter\x\fmtname xxxxxx\relax \def\y{splain}%
\ifx\x\y   
\gdef\SetFigFont#1#2#3{%
  \ifnum #1<17\tiny\else \ifnum #1<20\small\else
  \ifnum #1<24\normalsize\else \ifnum #1<29\large\else
  \ifnum #1<34\Large\else \ifnum #1<41\LARGE\else
     \huge\fi\fi\fi\fi\fi\fi
  \csname #3\endcsname}%
\else
\gdef\SetFigFont#1#2#3{\begingroup
  \count@#1\relax \ifnum 25<\count@\count@25\fi
  \def\x{\endgroup\@setsize\SetFigFont{#2pt}}%
  \expandafter\x
    \csname \romannumeral\the\count@ pt\expandafter\endcsname
    \csname @\romannumeral\the\count@ pt\endcsname
  \csname #3\endcsname}%
\fi
\fi\endgroup
\begin{picture}(252,184)(44,636)
\put(102,724){\makebox(0,0)[lb]{\smash{\SetFigFont{10}{12.0}{rm}
$\downarrow$}}}
\put( 65,715){\makebox(0,0)[lb]{\smash{\SetFigFont{17}{20.4}{rm}EA}}}
\put(107,736){\makebox(0,0)[lb]{\smash{\SetFigFont{10}{12.0}{rm}$\tau$}}}
\put(174,770){\makebox(0,0)[lb]{\smash{\SetFigFont{10}{12.0}{rm}
$y_1\longrightarrow$}}}
\put(127,680)  
{\makebox(0,0)[lb]{\smash{\SetFigFont{14}{16.8}{rm}
SG {\em III}}}}
\put(132,655)  
{\makebox(0,0)[lb]{\smash{\SetFigFont{10}{16.8}{rm}
1RSB}}}
\put(185,655)
{\makebox(0,0)[lb]{\smash{\SetFigFont{10}{16.8}{rm}
SG {\em IV}}}}
\put(120,799){\makebox(0,0)[lb]{\smash{\SetFigFont{17}{20.4}{rm}PM}}}
\put(124,766){\makebox(0,0)[lb]{\smash{\SetFigFont{10}{12.0}{rm}0}}}
\put(229,640){\makebox(0,0)[lb]{\smash{\SetFigFont{12}{14.4}{rm}
$y_4=10 |y_6|$}}}
\put(191,715){\makebox(0,0)[lb]{\smash{\SetFigFont{17}{20.4}{rm}SG}}}
\put(115,640)
{\makebox(0,0)[lb]{\smash{\SetFigFont{12}{14.4}{rm}
$y_4=|y_6|$}}}
\end{picture}
\vspace{.3cm}
\caption{$y_1-\tau$ phase diagram for $w>0$, $ y_6<0$.
In the SG{\em III} phase $q(x)$ is as in figure 2c.
Dynamically this phase splits up in a 1RSB phase and a SG {\rm IV} phase
, see fig. 2b,d. }
\label{fig5}
\end{figure}

\begin{figure}[h]
\hspace{0.25 \hsize}
\epsfxsize=0.45\hsize
\epsffile{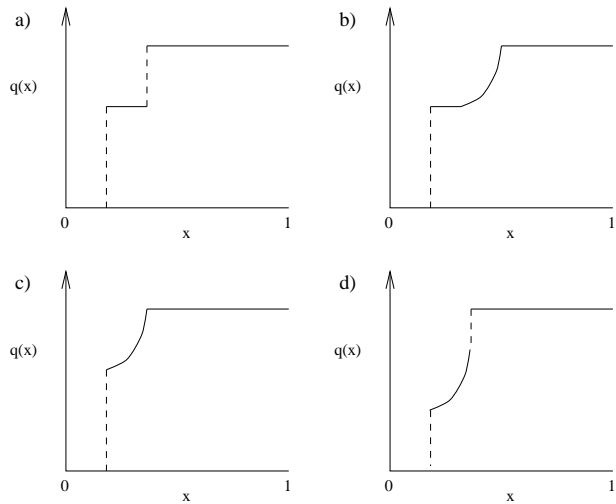}
\caption{Some unsuccessful attempts}
\label{fig6}
\end{figure}

\section{Summary}

We have presented the details of a new description for phase transitions in
site-disordered magnets, a field that was started by us in 1992.
As described in a letter of 1993~\cite{NEPL},
the approach starts from considering
an infinity of macroscopic order parameters. This approach is
different from, but  probably
equivalent to, the space-dependent order parameter field showing
up usually, such as in the model of section
\ref{sitepaironly}.

For the case of a random-site Ising spin glass, the central result
is given by eqs. (\ref{bFVBc}), (\ref{Psi=}), (\ref{Phn=}), and
(\ref{Xl=}). Simplifications are found in the combined limit of
small concentrations and large  coordination number $z$.
Here the random site problem generalizes the
mean field model with random bonds, the spin glass phase of the
Hopfield model and of the SK model.
Only the pair overlap order parameter
 $q_{\alpha\beta}$ plays a role.
In general, an infinity of multistate overlaps
$q^{(\ell)}_{\alpha_1\cdots\alpha_k}$, and conjugated overlaps
 $p^{(\ell)}_{\alpha_1\cdots\alpha_k}$ occurs.
A resummation has been done, which is useful for low temperatures.
This approach simplifies a bit for small concentrations, where
a close relation is found to the spin glass
model with diluted random bonds of Viana and Bray.
In this regime the random site
model exhibits a percolation threshold for short range
systems and the ``concentration scaling'' $T_g\sim c$
for RKKY systems.

We have generalized the approach to the case of vector spins.
This has led to a criterion for the transition temperature at
low concentrations.

In dilute YGd single crystals, closely related to
dilute  AuFe and CuMn crystals,
a thermodynamic transition to a state with ``complete''
spin density wave formation occurs~\cite{Wenger}.
The formation of incomplete
spin density waves in metallic spin glasses have been observed
by neutron scattering~\cite{Werner}.
Therefore incomplete spin density waves have been presumed
to be the cause of spinglass behavior in metals~\cite{Werner2}.
In our theory we indeed see that the effective coupling
$J_{\rm eff}(k)=\hat J(k)/(1-c\beta \hat J(k)(q_d-q_{EA}))$
enhances the quantitative contribution of
maxima in $\hat J(k)$. Nevertheless, thermodynamics
involves all wavevectors (see, e.g. (\ref{p=kint}))
without decisive role for
the special wavevectors connected to spin density waves.
As expected already by many, and confirmed by the experiments
of Weissman~\cite{Weissman},
incomplete spin density waves do not play a distinctive
role in the spin glass phase.
Although our present theory should be capable to explain why dilute
YGd has spin density waves, while AuFe and CuMn have a spin glass
phase, it is not clear to us how this should be shown.
Let us recall that in the second order cumulant expansion
(Gaussian approximation) the spin glass transition always
precedes spin density  waves or ferromagnetism~\cite{NEPL}.

Our present theoretical analysis also applies to the
ill understood cluster glass or mictomagnetic phase.
In metallic systems this is the phase observed
between the low
concentration spinglass and high concentration (anti-)
ferromagnet, see \cite{Mydoshboek} for a review.
We have considered the Ginzburg-Landau expansion
for our site-disordered field theory, within the
second order cumulant expansion (Gaussian approximation
in replica space).
These expressions show that the prefactors of the
cubic and quartic terms in the Ginzburg-Landau
expansion may go through zero at certain points.
The physically interesting cases are when either
the prefactor of the cubic term vanishes, or the one
of the quartic term that is responsible for
replica symmetry breaking. We show that in the first
case there is a transition to a phase with one step of
replica symmetry breaking, while in the latter there may
occur new spin glass phases between the standard spin glass
and the replica symmetric Edwards-Anderson phase.

It would be interesting to analyze the clusterglass
experimentally in more detail, and in particular to look for
new phases.
Theoretically it is worth to analyze loop effects for the
new phases. They may shed insight on the question whether fluctuations
(de-) stabilize the new phases, though their effect may also
be a mere shift of the transition point.
This problem is partly related to the
notorious problem of the renormalization of the AT-line.~\cite{BRob}
Finally the question of reentrant phases should also fall within
the scope of the theory presented here.

\acknowledgments
Throughout the years
the authors have benefited from discussions with
Jean-Philippe Bouchaud,
Ton Coolen, Cyrano de Dominicis, Jean-Marc Luck,
John Mydosh, Giorgio Parisi, Henri Orland, David Sherrington,
and many others.

\end{document}